\documentclass[aps,prx,twocolumn,superscriptaddress]{revtex4}
\usepackage{graphicx,amssymb,amsmath,amsfonts,empheq}
\usepackage[dvipsnames]{xcolor}
\usepackage{color}
\usepackage{braket}
\usepackage{latexsym}
\usepackage{upgreek}
\usepackage{dsfont}
\usepackage{bm}
\usepackage{multirow}
\usepackage{enumitem}
\usepackage[normalem]{ulem}
\usepackage{url}
\usepackage{hyperref}
\hypersetup{
colorlinks = true,
linkcolor=magenta,
citecolor=[rgb]{0.15,0.65,0.13},
urlcolor = [rgb]{0.25, 0.41, 0.88}}
\usepackage{bbm}

\newcommand{\Tr}{\mathrm{Tr}}

\newcommand{\mO}{\mathcal{O}}

\begin{document}
\title{Measurement-Driven Phase Transition within a Volume-Law Entangled Phase}

\author{Sagar Vijay}
\affiliation{Department of Physics, Harvard University, Cambridge, MA 02318, USA}
\begin{abstract}
We identify a phase transition between two kinds of volume-law entangled phases in non-local but few-body unitary dynamics with local projective measurements.  In one phase, a finite fraction of the system belongs to a fully-entangled state, one for which no subsystem is in a pure state, while in the second phase, the steady-state is a product state over extensively many, finite subsystems.   We study this ``separability" transition in a family of solvable models in which we analytically determine the  transition point, the evolution of certain entanglement properties of interest, and relate this to a mean-field percolation transition.  Since the entanglement entropy density does not distinguish these phases, we introduce the \emph{entangling power} -- which measures whether local measurements outside of two finite subsystems can boost their mutual information -- as an order parameter, after considering its behavior in tensor network states, and numerically studying its behavior in a model of Clifford dynamics with measurements.  We argue that in our models, the separability transition coincides with a transition in the computational ``hardness" of classically determining the output probability distribution for the steady-state in a certain basis of product states. A prediction for this distribution, which is accurate in the separable phase, and should deviate from the true distribution in the fully-entangled phase, provides a possible benchmarking task for quantum computers. % We comment on the possibility of this transition in other settings.
\end{abstract}
%\date{\today}
\maketitle

{\bf \emph{Introduction:}} New phases of entangled, out-of-equilibrium quantum matter have been an exciting topic of recent study.  A sufficiently generic, isolated quantum system is expected to approach local, thermal equilibrium under its own unitary dynamics, develop an extensive amount of entanglement entropy \cite{Srednicki1993Entropy, Srednicki1994Chaos, Deutsch1991Quantum}, and lose local memory of its initial state.  Random, local, unitary quantum circuits have provided a useful theoretical toolbox to study this delocalization of information and development of ``volume-law" entanglement on the  approach to equilibrium \cite{Nahum2017Quantum,Zhou2018Emergent,Keyserlingk2018Operator,Nahum2018Operator,Nahum2018Dynamics,Vijay2018Finite-Temperature,You2018Entanglement,Rakovszky2018Diffusive,Chan2018Spectral,Chan2018Solution,Bertini2019ESMMMMQC,Rakovszky2019Entanglement}.

Recent studies have revealed that in the presence of both unitary dynamics and local, projective measurements, there is a phase transition between a steady-state with volume-law or area-law scaling of the von Neumann entanglement entropy as the measurement rate is increased  \cite{Skinner2019MPTDE, Li2018QZEMET, Li2019METHQC,Bao2019TPTRUCWM,Jian2019MCRQC,Choi2019QECEPTRUCWPM,Gullans2019Dynamical,Zabalo2019Critical,Gullans2019Probes,Szyniszewski2019ETFVWM,Tang2019MPTCSNMDRGC,napp2019efficient}.  This volume-law state, however, appears qualitatively different than thermal quantum matter. In one spatial dimension, this volume-law state is characterized by a universal, logarithmic correction to the entanglement entropy \cite{Li2019METHQC, fan2020selforganized}, and the stability of this phase in the presence of measurements has been related to the theory of quantum error-correcting codes \cite{fan2020selforganized, Gullans2019Dynamical, Choi2019QECEPTRUCWPM}.  
%More recent work has shown that measurements in a fixed basis can lead to the emergence of area-law-entangled phases with interesting quantum orders \cite{sang2020measurement,lavasani2020measurementinduced}, reminiscent of the stabilization of topological phases in measurement-based quantum computation \cite{dennis2002topological}.

The local thermalization of a quantum system, however, can potentially be consistent with different kinds of multi-partite patterns of entanglement in the state, that can nevertheless be distinguished by a series of local operations.  In this work, we identify one such distinction by identifying a phase transition between two \emph{volume-law entangled} phases that emerge in the dynamics of a state undergoing non-local, but few-body, unitary evolution and local projective measurements.  In one phase, a finite fraction of the system belongs to a \emph{fully-entangled} state, one for which the entanglement entropy is non-zero for any bipartitioning;  in the other phase, the full state of the system is given by a product state over extensively many, finite subsystems.  A caricature of this \emph{separable} phase is a random configuration of Bell pairs, for which the typical entanglement entropy density of a subsystem is non-zero, even though the system is globally in a relatively simple product state.   

We present a family of solvable models of these dynamics in which both this phase transition, and certain entanglement properties of the evolving state can be studied exactly.  The solvability of our models stems from the fact that these few-body unitary dynamics with projective measurements can be re-cast as a measurement-outcome-dependent unitary transformation on the initial state; a graphical description of this unitary transition leads to a precise connection between this phase transition and mean-field percolation. This is reminiscent of connections between two-dimensional classical percolation, and the measurement-driven entanglement transition in one spatial dimension \cite{Skinner2019MPTDE, Jian2019MCRQC,Bao2019TPTRUCWM}, though the techniques used here differ substantially, as the unitary gates in our  models constitute ``instantaneous quantum polynomial-time" (IQP) dynamics \cite{IQP_Hard} which are not in the Clifford group, and for which we do not invoke a ``minimal-cut" prescription to compute the entanglement properties of interest.  Our techniques also allow us to show that the fully-entangled phase attained in the presence of measurements differs substantially from a random volume-law-entangled (Page) state \cite{Page:1993fv}, as the steady-state in an $N$-site system only requires $O(N)$ few-body but non-local unitary gates to be prepared from a product state.%We verify our predictions by simulating a restricted instance of these dynamics which are in the Clifford group. 

Since the entanglement entropy density alone does not distinguish between the fully-entangled and separable phases, we identify a new order parameter for this phase transition, termed the \emph{entangling power} of the state, which is defined as the change in the quantum mutual information between two finite subsystems after performing projective measurements in the remainder of the system.  Qualitatively, the entangling power probes the extent to which measurements can leverage the entanglement in the steady-state to communicate information between subsystems.  We argue that this is an order parameter for the transition by considering its behavior in a random tensor-network state, and through numerical simulations of Clifford dynamics with measurements, from which we extract critical exponents for the behavior of this quantity near the transition point.

We conclude by identifying a transition in the classical hardness of determining the probability distribution of the steady-state over a particular basis of product states that coincides with the separability transition in our toy model. This ``computational phase transition" is related to the fact that sampling the distribution of measurement outcomes in a certain basis for a state produced by a sufficiently random IQP unitary circuit requires calculating a ``partition function" for an Ising model with complex weights, for which there is no known classically efficient algorithm \cite{IQP_Hard}.  Nevertheless, we argue that the structure of the wavefunction in the separable phase permits a classically efficient way to perform this task, and that this breaks down in the fully-entangled phase.  The transition in the hardness of this sampling task provides an intriguing test for noisy intermediate-scale quantum computers (NISQ) \cite{Preskill_2018}, which have already demonstrated the ability to generate fully-entangled qubit states \cite{Full_Ent_1, Full_Ent_2}, and can sample the outputs of random quantum circuits \cite{arute2019quantum}.
 
A number of interesting questions remain to be studied about these dynamics, which we leave for future work.  First, it would be interesting to determine if the transition between fully-entangled and separable phases can occur within the volume-law entangled phase of spatially local unitary dynamics with projective measurements; this may generically be possible, and complementary to the already observed measurement-driven entanglement transition. Second, the universality class of this phase transition may differ from infinite-dimensional percolation in a more generic setting, where the wavefunction does not admit a graphical description. In addition, it would be interesting to see if the classical difficulty in simulating the quantum dynamics of these two volume-law entangled phases is more universally distinct, outside of the concrete proposal for our models, and if the fully-entangled phase could potentially provide a useful source of quantum error-correcting codes with low-weight ``parity-check" operations; we leave this for future study. 

During completion of this work, we became aware of a new version of Ref. \cite{Gullans2019Dynamical}, which includes a brief discussion of a similar transition in non-local unitary dynamics with projective measurements, by invoking a minimal-cut prescription for the calculation of the entanglement.  

{\bf \emph{Unitary Dynamics with Measurements:}} We consider the evolution of $N$ spins that are initialized in a product state  in the Pauli-$X$ basis.  At each timestep, we apply a two-site unitary gate -- which are diagonal in the Pauli-$Z$ basis -- between a randomly chosen pair of spins.   While these unitary operations mutually commute, they generate a high degree of entanglement when acting on states in a generic basis.  %The most generic form of such a two-site gate is $\sum_{s_{1},s_{2}\in\{0,1\}} \exp[i\theta s_{1}s_{2} + v_{1}s_{1} + v_{2}s_{2}]\ket{s_{1},s_{2}}\bra{s_{1}, s_{2}}$ up to an overall phase; 
After applying these gates, we measure a randomly-chosen spin in the $Z$ basis; this spin is then rotated back into the $X$ basis to the state $\ket{\rightarrow}$, so that it can continue to entangle with other spins.  A single ``timestep" of our dynamics consists of applying $\Gamma_{u}$ two-site unitary operations and $\Gamma_{m}$ measurements.  As shown in the Supplemental Material, the evolving wavefunction for these dynamics $\Psi(\boldsymbol{s};t)\equiv \langle s_{1},\ldots,s_{N}|\Psi(t)\rangle$ with $s_{j} \equiv (1-Z_{j})/2$ (so that $s_{j} = 0$ or $1$ if the spin $j$ points up or down, respectively) %(e.g. the state $\ket{11}$ has both spins pointing down), 
always takes the form
\begin{align}\label{eq:graph_evolution}
\Psi(\boldsymbol{s};t) = \frac{1}{\sqrt{D}}\exp\left[{ \frac{i}{2} \boldsymbol{s}^{T}\boldsymbol{\theta}(t)\boldsymbol{s} + i\boldsymbol{w}(t)\cdot \boldsymbol{s}}\right]
\end{align}
where $\boldsymbol{\theta}(t)$ is an $N\times N$ symmetric matrix, $\boldsymbol{w}(t)$ is an $N$-component vector.   Both of these quantities are defined modulo $2\pi$.  Additionally, $D = 2^{N}$ is the Hilbert space dimension of the system. In other words, the dynamics with measurements that we have described can be re-cast as a measurement-outcome-dependent unitary transformation on the initial state, which we can readily compute. Furthermore, these dynamics encompass two distinct types of evolution. %The dynamical rules for the evolution of these quantities can be determined.  Notably, the evolution of $\boldsymbol{\theta}(t)$, and consequentially the entanglement dynamics of the state, are independent of the measurement outcomes. %We now discuss the rules for the evolution of this matrix, as we apply unitary operations and measurements.  %While $\boldsymbol{\theta}(t)$ is naturally thought of as an adjacency matrix for a weighted, undirected graph with $N$ nodes, 

\begin{figure}[t]
$\begin{array}{c}
\includegraphics[trim = 0 0 0 0, clip = true, width=0.31\textwidth, angle = 0.]{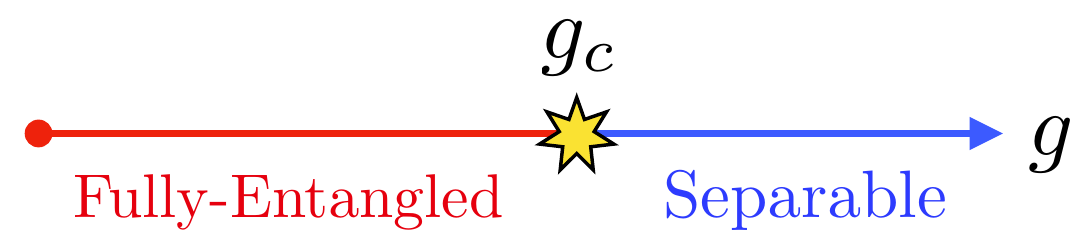}\\
\text{(a)}\\\\\\
\includegraphics[trim = 0 0 0 0, clip = true, width=0.37\textwidth, angle = 0.]{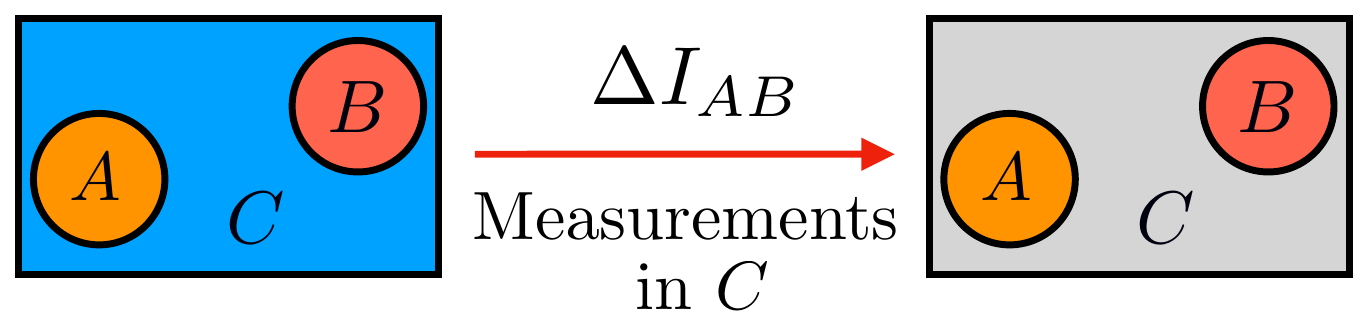}\\
\text{(b)}
\end{array}$
\caption{{\bf Phase Diagram:} As the measurement rate $\Gamma_{m}$ is increased relative to the rate of applied unitary gates $\Gamma_{u}$, as quantified by $g\equiv\Gamma_{m}/2\Gamma_{u}$, the wavefunction undergoes a sharp transition between a phase where a fraction of the system is in a fully-entangled state, and a phase where the steady-state resembles a separable, product state over small subsystems.  We propose an order parameter (b) as the change in the mutual information between two subsystems after performing measurements in the remainder of the system.}
  \label{fig:Phase_Diagram}
\end{figure}

\emph{Clifford Dynamics:} If the two-site unitary operations are exclusively control-$Z$ gates, which acts as ${CZ_{12}}\ket{s_{1}s_{2}} = (-1)^{s_{1}s_{2}}\ket{s_{1}s_{2}}$, these dynamics are classically efficiently simulable \cite{Gottesman_Knill}.  %Wavefunctions generated by acting with these gates on a state in the Pauli-$X$ basis have been studied in the context of measurement-based quantum computation \cite{Raussendorf, GraphStateEE}.  
In this case,  $\boldsymbol{\theta}(t) = \pi\boldsymbol{G}(t)$, where $\boldsymbol{G}(t)$ is a binary matrix with entries zero or one.  For a bi-partitioning of the system into sub-systems $A$ and its complement $\bar{A}$, the von Neumann entanglement entropy is known \cite{GraphStateEE2} to be 
$S_{A}(t) =\Tr[\rho_{A}(t)\log_{2}\rho_{A}(t)] = \mathrm{rank}_{\,\mathbb{F}_{2}}\big[ G_{A\bar{A}}(t) \big]$, where $G_{A\bar{A}}$ the off-diagonal part of the matrix $\boldsymbol{G}$, which connects sites in the $A$ and $\bar{A}$ subsystems.    
This matrix also appears in the natural choice of Pauli operators (``stabilizers'')  that fix the evolving state of the spins as $\mathcal{O}_{n}(t) |\Psi(t) \rangle = |\Psi(t) \rangle$ (for $n = 1,\ldots, N$); these operators are given up to an overall sign as $\mathcal{O}_{n}(t) \sim X_{n} \prod_{m=1}^{N} (Z_{m})^{\boldsymbol{G}_{nm}(t)}$.  Clifford states of this form have been studied in the context of measurement-based quantum computation \cite{Raussendorf}, and it is known that any stabilizer state can be brought into a state of this form through the action of local Clifford operations \cite{Stabilizer_GraphState_Equivalence1, Stabilizer_GraphState_Equivalence2}.

\emph{Non-Clifford Dynamics:}  Unitary dynamics consisting of more generic, mutually-commuting two-qubit gates which  generate no entanglement in the Pauli-$Z$ basis are often referred to as ``instantaneous quantum polynomial-time" (IQP) dynamics, and the output distribution of measurements in a generic basis from a state generated by a random IQP circuit classically intractable \cite{IQP_Hard, IQP_Hard2}.  We will return to the tractability of sampling from an IQP circuit in the final section of this work.  %First, rule (1) above remains unchanged; when a $Z$ measurement and rotation into the Pauli-$X$ basis are performed on spin $k$, then the node $k$ is disconnected from all nodes in the graph described by $\boldsymbol{\theta}(t)$.  We note that this dynamical rule is independent of the measurement outcome. The second rule is modified so .  Nevertheless, our object of interest will be the {binary} matrix $\boldsymbol{G}(t) \equiv  || \boldsymbol{\theta}(t) ||$, in which each entry is one iff the corresponding entry in $\boldsymbol{\theta}$ is non-zero (mod $2\pi$).  Subsystem $A$ is disentangled from the system if and only if $G_{A\bar{A}}$ -- the off-diagonal part of $\boldsymbol{G}(t)$ -- is the zero matrix.  We argue that in a thermodynamically large system, the dynamics of this binary matrix are the same, across both Clifford and non-Clifford dynamics, resulting in an identical phase transition in the separability of the wavefunction. 
%Mutually commuting unitary gates, which generate no entanglement in the Pauli $Z$ basis, are referred to as IQP circuits; sampling .

For both of these kinds of dynamics, we would like to determine when a given subsystem is disentangled from the remainder of the system.  As we show in the Supplemental Material, a subsystem $A$ is disentangled from its complement $\bar{A}$ if and only if $\theta_{A\bar{A}}$ -- the off-diagonal part of $\boldsymbol{\theta}(t)$ that connects nodes in sub-systems $A$ and $\bar{A}$ -- is the zero matrix.   Equivalently, we may restrict our attention to an $N\times N$ binary matrix $\boldsymbol{G}(t)$, defined so that each entry is one if the corresponding entry in $\boldsymbol{\theta}(t)$ is non-zero, and is zero otherwise.  This matrix is naturally thought of as an adjacency matrix for an undirected graph on $N$ nodes, one for each spin in the system. If the two-site unitary gates consist of exclusively control-$Z$ gates, then this graph describes the evolution of the stabilizers that define this state, as discussed previously.  The dynamics of this graph, however, are the same for any choice of commuting unitary gates, and are given by the following dynamical rules
\begin{enumerate} 
\item When a measurement is performed in the $Z$ basis on spin $k$, and the spin is then rotated into the Pauli $X$ basis, the graph evolves by removing all of the bonds from node $k$. 
\item When a two-site unitary gate is applied, the corresponding nodes in the graph are connected by bond. %if none is present, and a bond is removed if the spins are already connected.
%\item If a measurement in the $Z$ basis yields the result that the spin is pointing down, then the filling of all nodes that were previously connected to node $k$ are changed.  
\end{enumerate} 
The first rule is exact, while the second rule becomes exact in the thermodynamic limit.  For the latter rule to be valid for a given pair of spins, either one has to have undergone a measurement more recently than that exact pair has been acted upon by a unitary gate, which is exactly true in the thermodynamic ($N\rightarrow\infty$) limit. %Since the number of pairs of sites is a given spin undergoes a measurement, as per the first rule, before that same pair of spins is acted upon by a two-site unitary gate.  

%The matrix $\boldsymbol{\theta}(t)$ is naturally thought of as an adjacency matrix for a weighted, undirected graph with $N$ nodes.  Our object of interest will be the \emph{binary} matrix $\boldsymbol{G}(t) \equiv  \boldsymbol{\theta}(t) \mod 2\pi$, in which each entry is one iff the corresponding entry in $\boldsymbol{\theta}$ is non-zero (mod $2\pi$).  This quantity naturally appears i

%We note that the entanglement properties of the wavefunction are encoded in the adjacency matrix $G(t)$ \cite{GraphStateEE,GraphStateEE2}.  For a bi-partitioning of the system into sub-systems $A$ and $B$, %the adjacency matrix $G$ associated with this graph state may be written in the block form
%\begin{align}
%G(t) = \left(\begin{array}{cc} G_{AA}(t) & G_{AB}(t)\\
%G_{AB}^{T}(t) & G_{BB}(t)
%\end{array}\right)
%\end{align}
%the von Neumann entanglement entropy $S_{A}(t)$, given our initial state is   $S_{A}(t) =\Tr[\rho_{A}(t)\log_{2}\rho_{A}(t)] = \mathrm{rank}_{\,\mathbb{F}_{2}}\big[ G_{AB}(t) \big]$, where $G_{AB}$ is the off-diagonal part of the adjacency matrix $G$, which describes how the nodes in sub-systems $A$ and $B$ are connected \cite{GraphStateEE}.
%We prove this formula in the appendix.  

{\bf \emph{``Separability" Phase Transition:}} The steady-state entanglement properties of the wavefunction change dramatically as the measurement rate is tuned.  %In the presence of both unitary dynamics and projective measurements, the wavefunction of the system may always be written as a tensor product over fully-entangled states of {clusters} of spins.  Schematically,
%\begin{align}
%|\Psi(t)\rangle \sim \bigotimes_{m}|\varphi_{m}(t)\rangle
%\end{align}
%where each state $|\varphi_{m}(t)\rangle$ of a fraction of the spins is entangled for any bipartitioning. 
When the measurement rate is sufficiently small, a finite fraction of the spins will be \emph{fully entangled}, and will belong to a single non-separable state, one for which the von Neumann entanglement entropy is non-zero for any bipartitioning.  As the measurement rate is increased, however, there will eventually be a sharp phase transition, beyond which the wavefunction will resemble a product state over small clusters of spins.  This {separability} phase transition is summarized in Fig. \ref{fig:Phase_Diagram}.  The intuition for this result comes from the graphical rules for the evolution of the separability of state, which involves the growth and fragmentation of a graph due to the unitary gates and measurements. % -- and from the fact that the von Neumann entanglement entropy of a bipartition is zero only if there are no bonds that connect any of the nodes in the region $A$ and its complement, in the graphical representation of the wavefunction.     
We will analytically access this transition and study its properties in the following section.
%Schematically,
%\begin{align}
%\lim_{t\rightarrow\infty}|\Psi(t)\rangle = |\varphi_{1}\rangle\otimes\cdots\otimes|\varphi_{n}\rangle
%\end{align}

\emph{Graphical Evolution of the State:} Using the rules for the evolution of the graph $\boldsymbol{G}(t)$, we now study the evolution of the separability of the wavefunction.   %To describe the dynamics of the wavefunction, it is convenient to study the stabilizers that describe evolving state of the spins. Since the wavefunction may always be represented as a graph state, a canonical choice of stabilizers is of the form given previously, where each stabilizer consists of a single Pauli-$X$ operator multiplied by some product of Pauli-$Z$ operators on other sites.  
%In this canonical basis of stabilizers, a stabilizer will have ``size" $k$ if the stabilizer is a product of $k$ Pauli operators.
Let  $s_{k}(t)$ be the number density of sites in the graph that have degree $(k-1)$.  We emphasize that if the unitary gates are chosen to be control-$Z$ gates and the dynamics are in the Clifford group, then $s_{k}(t)$ is precisely the number density of  stabilizers of weight $k$ at time $t$ in this canonical basis of stabilizers described previously. 

Since the total number of nodes in the graph is conserved $\sum_{k\ge 1}s_{k}(t) = 1$ at any point in the dynamics.  In a time-continuum limit, the number density of nodes in the graph of a certain degree satisfies a rate equation
\begin{align}\label{eq:stab_evolution}
\frac{ds_{k}}{dt} = \sum_{j}s_{j}\,\gamma_{j\rightarrow k}  - s_{k}\sum_{j}\gamma_{k\rightarrow j}
\end{align}
where $\gamma_{j\rightarrow k}$ is the rate at which nodes of size $j-1$ transform into nodes of degree $k-1$ under the dynamics.  This rate equation manifestly conserves the total number of nodes as required.  From the dynamical rules for the graphical evolution of the state of the system, we observe that the degree of a node grows due the two-site unitary operations. %; the effect of this unitary operator is to add a bond between the two nodes on which the unitary acts.
%, so that a size-$k$ stabilizer is created when the unitary acts on a site corresponding to a node in the graph with degree $k-1$. 
Furthermore, a measurement removes all bonds from the corresponding node in the graphical representation of the state.  Putting these two elements of the dynamics together yields the expression 
$\gamma_{j\rightarrow k} = 2\,\Gamma_{u} \delta_{j,k-1} + \Gamma_{m}[k\,\delta_{j,k+1} + \delta_{k,1}(1-\delta_{j,1})]$. 
The last term in this expression arises since every measurement creates a node with zero degree.  

\begin{figure}
\includegraphics[trim = 0 0 0 0, clip = true, width=0.32\textwidth, angle = 0.]{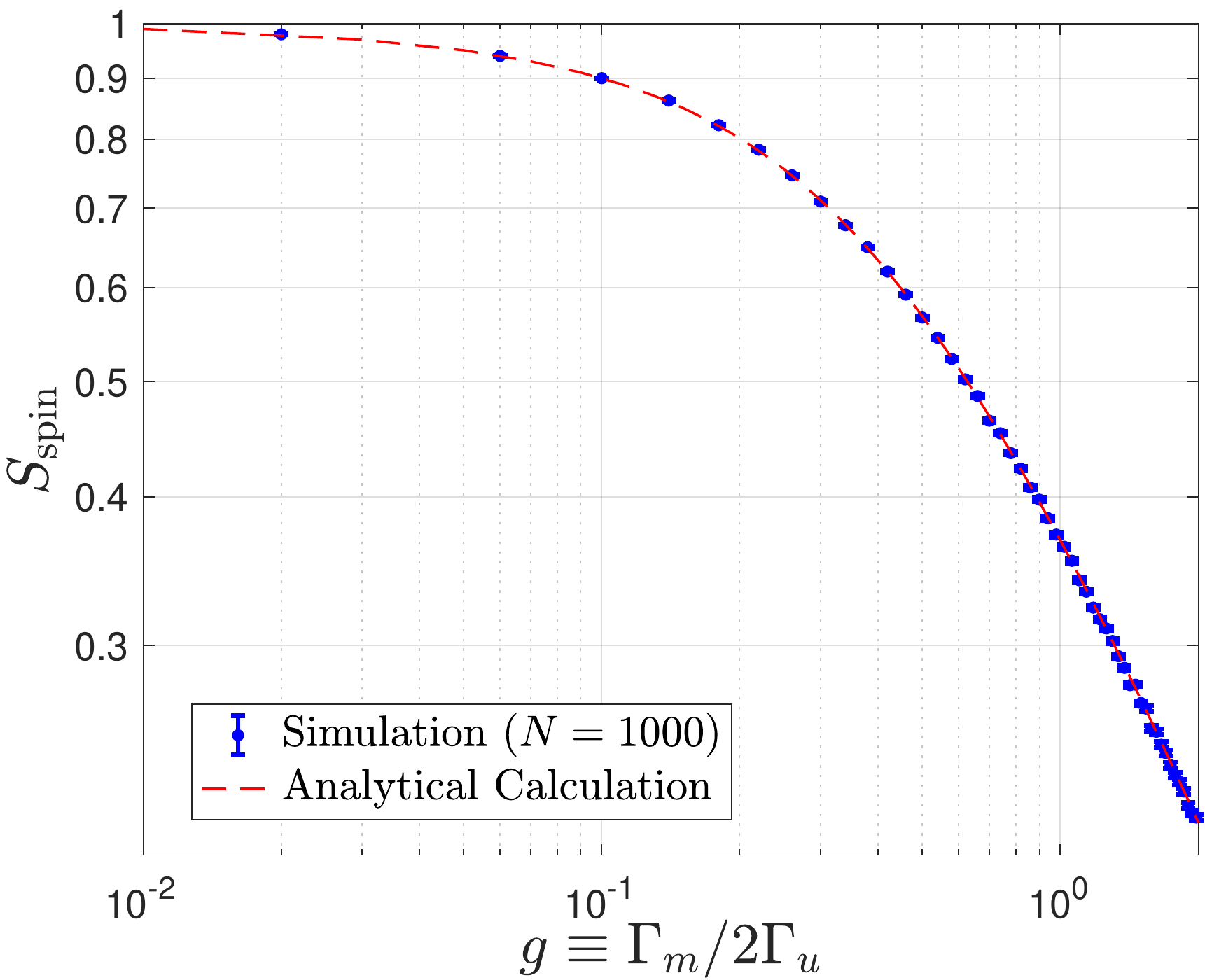}
\caption{{\bf Steady-State Entanglement Entropy of a Spin:} For the Clifford dynamics with measurements described in the text, the analytical result for the steady-state, von Neumann entanglement entropy of a spin Eq. (\ref{eq:steady_singlespin_ee}) matches with the entanglement, as calculated in numerical simulation of the dynamics for $N = 1000$ spins, shown in blue.}
  \label{fig:Spin_EE}
\end{figure}

The linearity of the rate equation permits an exact solution to Eq. (\ref{eq:stab_evolution}) in order to obtain the full, time-dependent dynamics of the graph, as shown in the Supplemental Material.  In the steady-state, we find that
\begin{align}\label{eq:steady_state_stab}
s_{k}^{(\infty)} = g\left[1 - \frac{\Gamma\left(k, 1/g\right)}{\Gamma(k)}\right]
\end{align} 
where $\Gamma(a,b)$ is the incomplete gamma function, and the dimensionless parameter $g$ is defined as $g \equiv {\Gamma_{m}}/({2\Gamma_{u}})$.  Eq. (\ref{eq:steady_state_stab}) is significant for a few reasons.  First, the average degree of a node is finite, and approaches $1/2g$ in the steady-state. Therefore, in the presence of these projective measurements, the number of unitary operations needed to prepare the steady-state from the initial state only scales as $O(g^{-1}N)$. %, in contrast to a naive expectation of $O(N^{2})$ two-qubit gates.  
Second, for Clifford dynamics, Eq. (\ref{eq:steady_state_stab}) is precisely the distribution of stabilizer sizes in the canonical basis.  As an illustration of the accuracy of this result, we observe that for Clifford dynamics, the von Neumann entanglement entropy of a single spin -- denoted $S_{\mathrm{spin}}(t)$ -- is precisely the fraction of stabilizers which are of length greater than one, so that $S_{\mathrm{spin}}(t) = 1 - s_{1}(t)$.  From the exact solution to Eq. (\ref{eq:stab_evolution}) derived in the Supplemental Material, this yields 
\begin{align}
S_{\mathrm{spin}}(t) = 1 - g + g \,e^{-(1-e^{-\Gamma_{m}t})/g}\left(1 - \frac{e^{-\Gamma_{m}t}}{g}\right)
\end{align}
so that in the steady-state, the average entanglement of a spin is 
\begin{align}\label{eq:steady_singlespin_ee}
\lim_{\Gamma_{m}t\rightarrow\infty}S_{\mathrm{spin}}(t) = 1 - g\left(1 - e^{-1/g}\right)
\end{align}
The equilibrium value of the average, single-spin entanglement entropy is plotted in Fig. \ref{fig:Spin_EE}, along with a calculation of the average entanglement entropy of a single spin in a full simulation of the Clifford dynamics (as described previously, involving control-$Z$ gates and projective measurements followed by single-qubit rotations acting on an initial  state in the Pauli $X$ basis) in a system of $N=1000$ spins.  For a general IQP circuit, the average entanglement of a spin is upper-bounded by $S_{\mathrm{spin}}(t)$. 

%\begin{align}
%\frac{dn_{s}}{d\tau} = &\frac{g}{2}\sum_{q\le s} q(s-q) n_{q} n_{s-q}- (1 +wg)s n_{s} \nonumber\\
%&+ (s+1)n_{s+1} + w\delta_{s,1}
%\end{align}
%Introducing the generating function 
%\begin{align}
%F(t) &\equiv |\langle\Psi(0)|\Psi(t)\rangle|^{2} = \frac{1}{D}\Big|\sum_{\boldsymbol{s}} \Psi(\boldsymbol{s};t)\Big|^{2} \nonumber\\
%&= \frac{1}{D^{2}}\sum_{\boldsymbol{r},\boldsymbol{s}} \prod_{n<m} (-1)^{G_{nm}(t)\left[s_{n}s_{m}+ r_{n}r_{m}\right]}
%\end{align}
%
%Averaging over the ensemble of graphs yields
%\begin{align}
%\overline{F(t)} = \frac{1}{D^{2}}\sum_{\boldsymbol{s},\boldsymbol{r}}e^{-H[\boldsymbol{s},\boldsymbol{r}]}
%\end{align}
%where $H[\boldsymbol{s},\boldsymbol{r}] = w\sum_{n,m}\left(2 s_{n}s_{m}r_{n}r_{m} - s_{n}s_{m} - r_{n}r_{m}\right)$.

{\bf \emph{Properties of the Phase Transition:}}
%The steady-state entanglement of the spins changes dramatically as the dimensionless parameter $g$ is tuned.  When $g$ is sufficiently small, a macroscopic number of spins will be a part of a \emph{fully entangled} state, one for which any bipartitioning has a non-zero von Neumann entanglement entropy.  In contrast, for sufficiently large $g$, the steady-state wavefunction will resemble a simple product state, for which any finite collection of spins will each typically be part of distinct, disentangled components of the full state of the system.  %Since the measurements have the effect of Therefore, the connectedness of the graph determines whether the wavefunction \cite{Molloy_Reed}.
Properties of the wavefunction around the separability phase transition may be determined by assuming that in the steady-state, the graph $\boldsymbol{G}$ is {random}, with node degrees drawn from the steady-state distribution (\ref{eq:steady_state_stab}).  The predictions that we obtain by employing this assumption are consistent with a mean-field percolation transition, and compare favorably with simulations of Clifford dynamics.

%Using this assumption about the steady-state, we now determine the point at which the separability phase transition occurs.  
First, the graph corresponding to the steady-state wavefunction contains a connected cluster of nodes encompassing a finite fraction of the system, and the corresponding wavefunction is fully entangled, under the following conditions.  The probability that a random node in the graph will have degree-$k$ is $Q_{k} = s_{k+1}^{(\infty)}$, while the probability distribution for the degree of a node's reached by following a random bond is $P_{k} = 2g k\,s_{k+1}^{(\infty)}$\footnote{By following a random link to a node, we obtain a randomly chosen neighbor, which will have the degree distribution $P_{k}\sim k \,Q_{k}$, as there are $k$ different ways that each such node could be reached; the pre-factor of $2g$ provides the correct normalization.}. %For a random graph state, this criterion implies that 
%\begin{align}\label{eq:MR}
%\sum_{k\ge 1} k^2 s_{k+1}^{(\infty)} \ge 2\sum_{k\ge 1} k s_{k+1}^{(\infty)}
%\end{align}
Therefore, the number of nodes that can be reached by starting at a random node and following the bonds of the graph for $m$ steps\,\footnote{Here, we assume the node is part of a cluster that contains no loops.  This is valid as long as the cluster is of a finite size, and in the thermodynamic $N\rightarrow\infty$ limit. } is $(\sum_{k}(k-1)P_{k})^{m}$.  As a result, a randomly chosen spin belongs to an entangled cluster of spins of size $\ell$ where %size $\xi$ of a connected cluster of nodes in the steady-state -- which is the size of a fully-entangled cluster of spins -- is then
\begin{align}\label{eq:clust_size}
\ell = \sum_{m}\left[\sum_{k}(k-1)P_{k}\right]^{m} = \frac{1}{1  - (g_{c}/g)}
\end{align}
with $g_{c} \equiv {2}/{3}$.  %Equivalently the probability that two randomly chosen sites belong to the same cluster is $\sim \ell/N$, which vanishes in the thermodynamic limit when $g >g_{c}$. 
%in order for the steady-state wavefunction to be \emph{fully entangled}; evaluating this inequality for our fixed-point stabilizer distribution yields the condition that $g \le g_{c}$ with
 We observe that $\ell$ is finite for $g>g_{c}$, but \emph{diverges} as $g\rightarrow g_{c}$, signalling the transition between a wavefunction that resembles a simple product state, to one that is fully entangled.  We may also analytically determine the fraction of sites $m(g)$ that belong to the largest entangled cluster of spins in the system when $g<g_{c}$, as we show in the Supplemental Material.  Fig. \ref{fig:Cluster_Mass} shows our analytic result in the thermodynamic limit, which compares favorably to numerical simulations of the full Clifford dynamics of the state for system sizes up to $N = 2000$ spins.  % For random graphs with a specified degree distribution, this criterion for graph connectedness (\ref{eq:clust_size}) is known to be rigorously true  \cite{Molloy_Reed}.

\begin{figure}
\includegraphics[trim = 0 0 0 0, clip = true, width=0.32\textwidth, angle = 0.]{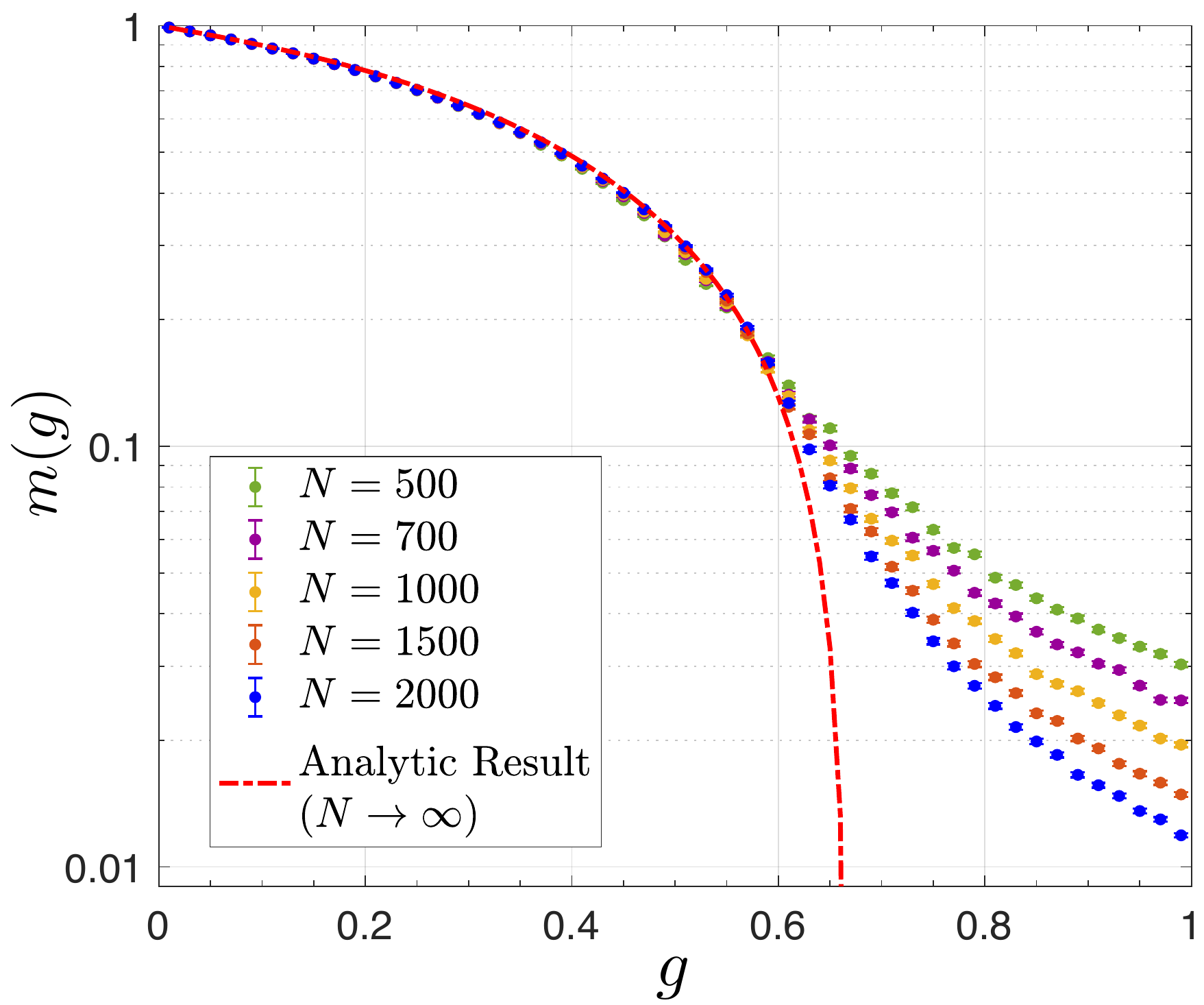}
\caption{{\bf The Largest, Fully-Entangled Cluster of Spins:} The relative size of the largest entangled cluster is shown in numerical simulations, for the indicated number of spins $N$.  The exact analytic result is obtained through the solution to an implicit equation for $m(g)$, derived in the Supplemental Material, which predicts a phase transition at $g_{c} = 2/3$.}
  \label{fig:Cluster_Mass}
\end{figure}

Other properties of the wavefunction near criticality may be determined by exploiting known results on mean-field percolation.  For the percolation transition, the number density of \emph{finite} clusters of size $k$, denoted $n_{k}$ is known to behave near the critical point as \cite{stauffer2018introduction} %random graphs that are constructed from nodes with degrees drawn from a fixed distribution. The study of such random graphs was initiated by Erd\"{o}s and R\'{e}nyi \cite{Erdos_Renyi}, who considered a uniform distribution over graphs with a fixed number of nodes and edges.  As the number of edges is increased, the Erd\"{o}s-R\'{e}nyi (ER) random graph undergoes a percolative phase transition which is in the same universality class as percolation in infinitely-many spatial dimensions.  This transition is characterized by the emergence of a connected cluster that encompasses a finite fraction of nodes.  The number density of \emph{finite} clusters of size $k$, denoted $n_{k}$, characterizes this phase transition, and is known to behave near the critical point as  
\begin{align}\label{eq:power_law}
n_{k}(g) \sim k^{-\tau} {e^{-k/s(g) }}
\end{align}
where $s(g) \sim |g-g_{c}|^{-1/\sigma}$ is the size of the largest finite cluster.  The universal exponents  $\sigma$, $\tau$ are given by $\sigma = 1/2$ and  $\tau = 5/2$ for mean-field percolation \cite{stauffer2018introduction}. %The power-law decay of the entangled cluster sizes is shown for the Clifford dynamics that we have simulated in Fig. \ref{fig:Cluster_Size}, which  shows a decay consistent with $n_{k} \sim k^{-\tau}$ near criticality. %The characteristic size of finite clusters diverges near criticality as ${\xi}(g) \sim |g_{c} - g|^{-1}$.
The first and second moments of this distribution (\ref{eq:power_law})
\begin{align}
m(g) \equiv 1 - \sum_{k} k\,n_{k}(g)\hspace{.2in}
\chi(g) \equiv \sum_{k}k^{2}n_{k}(g)\label{eq:fluctuations}
\end{align}
where the sum is over the {finite-sized} clusters in the graph \footnote{By definition, $\sum_{k} k \,n_{k} = 1$ in a finite system.} describe the fraction of spins that belong to the largest fully-entangled state, and the fluctuations in the cluster distribution, respectively.  A consequence of Eq. (\ref{eq:power_law}) is that in a finite-sized system, these quantities obey the scaling forms \cite{luczak1990component, bollobas1984evolution, stauffer2018introduction, borgs2001birth,Erdos_Renyi}
\begin{align}
m(g, N) &= N^{-1/3}\,f_{m}\left( N^{1/3}|g - g_{c}| \right)\\
\chi(g, N) &= N^{1/3}\,f_{\chi}\left( N^{1/3}|g - g_{c}| \right)\label{eq:fluctuations_scaling}
\end{align}
near criticality, with both $f_{m}(z)$ and $f_{\chi}(z)$, analytic functions of their arguments.  

\begin{figure}
\includegraphics[trim = 0 0 0 0, clip = true, width=0.34\textwidth, angle = 0.]{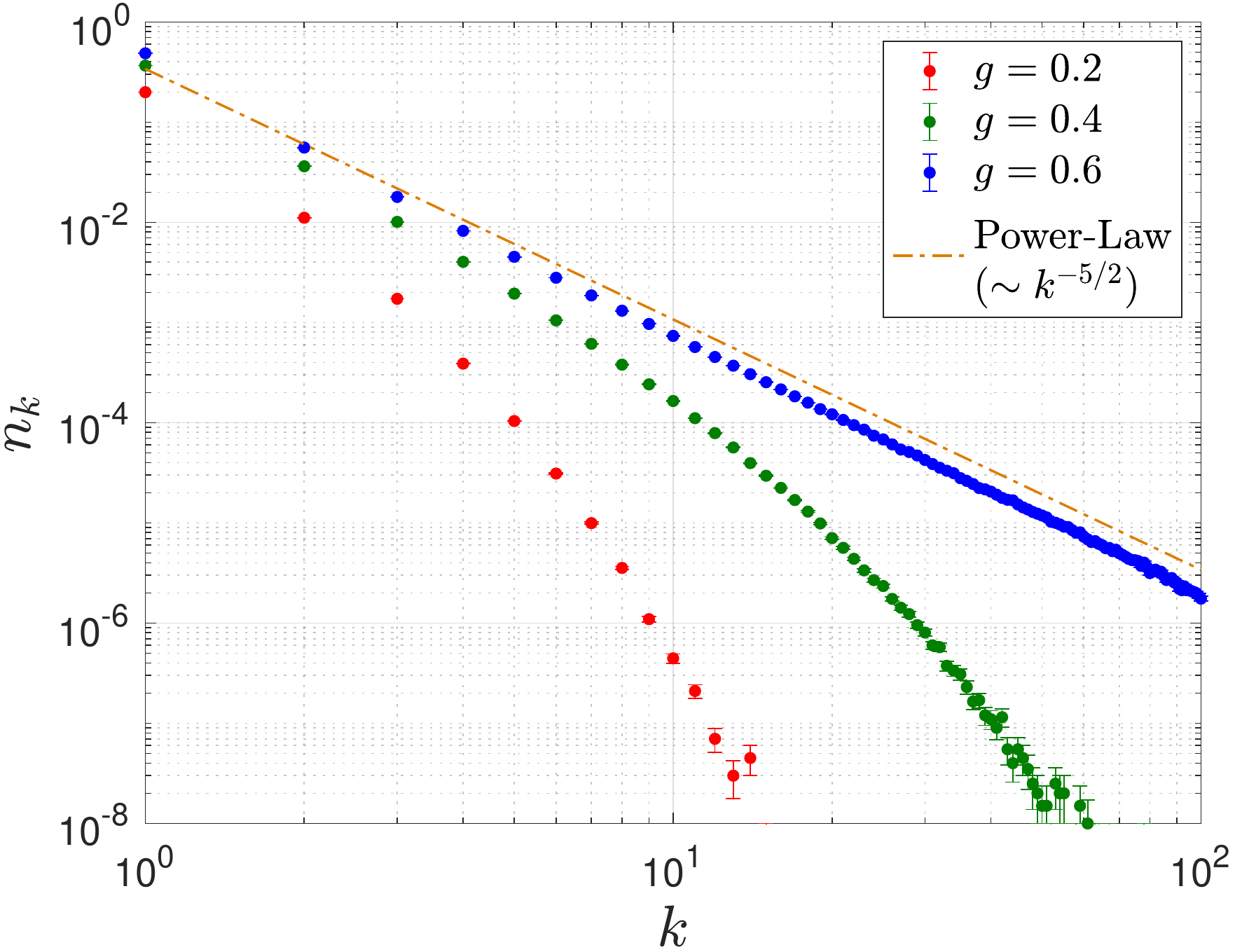}
\caption{{\bf Distribution of Entangled Cluster Sizes:}  We simulate the dynamics in a system of $N = 2000$ spins to obtain the steady-state.  By averaging over $10^{5}$ realizations of the dynamics for each value of $g$, we observe a power-law decay of $n_{k}\sim k^{-5/2}$ as criticality is approached.  }
  \label{fig:Cluster_Size}
\end{figure}

\begin{figure}
%$\begin{array}{c}
 \includegraphics[trim = 0 0 0 0, clip = true, width=0.36\textwidth, angle = 0.]{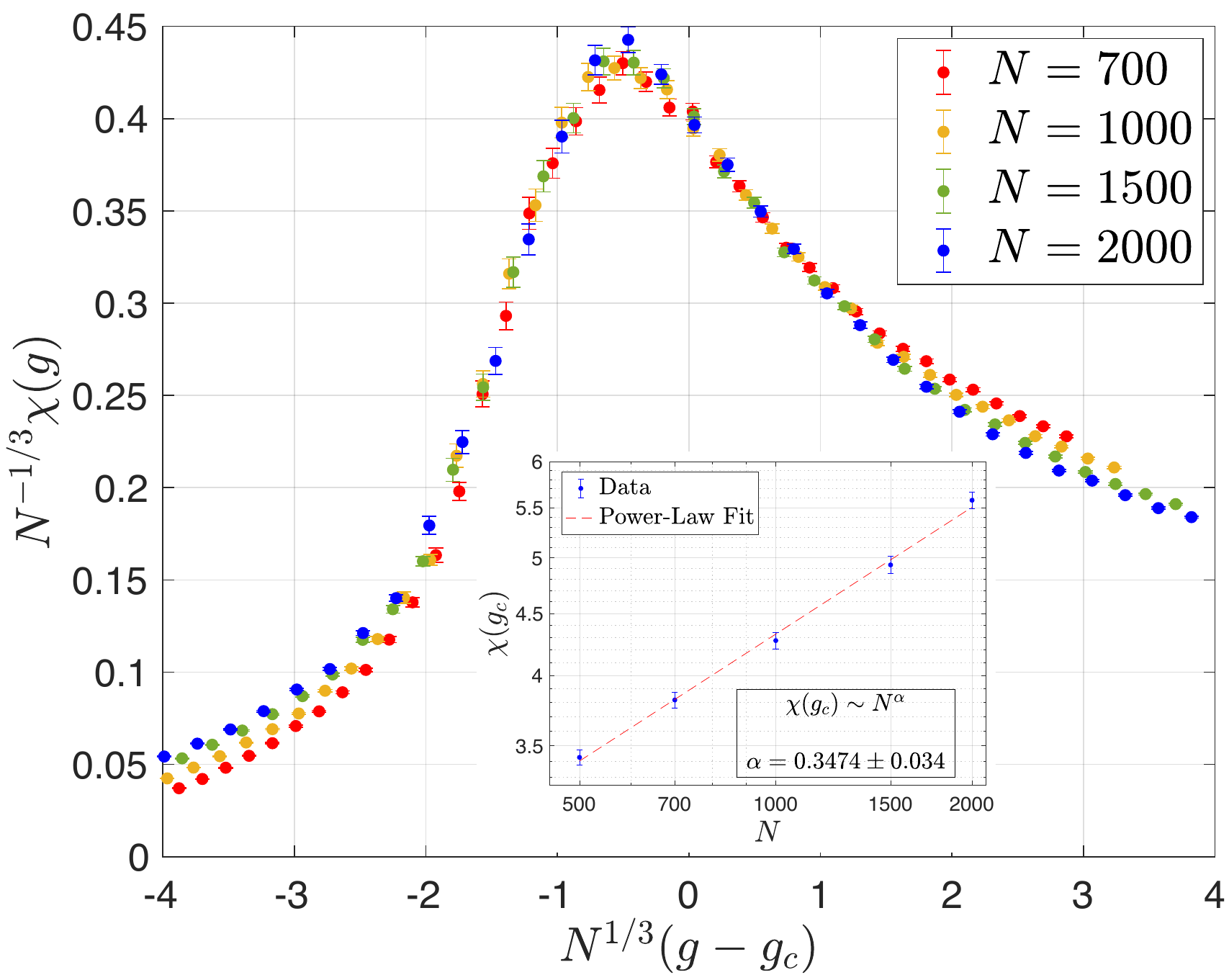}\\
 %\text{(a)}\\
%\includegraphics[trim = 0 0 0 0, clip = true, width=0.33\textwidth, angle = 0.]{chi_max_paper}\\
%\text{(b)}
%\end{array}$
\caption{{\bf Fluctuations in the size of entangled clusters near the transition:}  For the Clifford dynamics described in the text, we observe that the fluctuations in the sizes of the entangled clusters, as defined in Eq. (\ref{eq:fluctuations}) diverge near the critical point, and are consistent with the scaling form (\ref{eq:fluctuations_scaling}).}
  \label{fig:Cluster_Suscept}
\end{figure}

We verify the power-law decay of the number of fully-entangled clusters in Eq. (\ref{eq:power_law}), along with the scaling of the fluctuations in the size of these clusters in Eq. (\ref{eq:fluctuations_scaling}) in numerical simulations of Clifford dynamics.  The distribution of cluster sizes, shown in Fig. \ref{fig:Cluster_Size}, is obtained by simulating $10^{5}$ realizations of the dynamics for each value of $g$ in a system of $N = 2000$ spins.  The numerical results show the power-law decay of the cluster distribution as criticality is approached, which is consistent with the expected $n_{k}\sim k^{-5/2}$ behavior.  The scaling collapse in Fig. \ref{fig:Cluster_Suscept} is obtained similarly, by averaging the fluctuations in the cluster size over $10^{5}$ realizations of the dynamics for each indicated system size.  This collapse is consistent with the scaling form in Eq. (\ref{eq:fluctuations}).

{\bf\emph{An Order Parameter for the Transition:}} We now introduce a measurable order parameter that distinguishes between the fully-entangled and separable phases, and study its behavior near criticality.  To motivate our proposal, we observe that the mutual information between finite subsystems is small in both phases.  In the fully-entangled phase, this is because the density matrix for every finite region will be close to the identity (infinite temperature), while in the separable phase, the low mutual information is due to the fact that a typical group of spins will belong to a separable product state, and will not be entangled ``with each other"; a caricature of the steady-state in the separable phase is given by a random configuration of Bell pairs, for which the mutual information between a typical pair of spins is strictly zero.

To distinguish between these two forms of volume-law entanglement, we propose an order parameter -- termed the \emph{entangling power} of the state -- and study its behavior near criticality.  In the steady-state, we tri-partition our system into disjoint sub-systems ($A$, $B$, $C$), with $A$ and $B$ each consisting of a finite number of spins.  We now consider the change in the von Neumann mutual information between $A$ and $B$ after performing single-qubit measurements in the $C$ subsystem in an arbitrary basis.  We define the order parameter as this change in the mutual information, averaged over realizations of the dynamics, which we denote as $\overline{\Delta I_{AB}}$.  The entangling power probes the extent to which  measurements alone can {increase} the mutual information between qubits, by leveraging the existing entanglement in the steady-state.

The entangling power behaves as a natural order parameter for this transition.  First, consider a state $\ket{\Psi}$ for which two finite subsystems $A$ and $B$ are disentangled from each other; that is, the full state of the system can be written as $\ket{\Psi} = \ket{\psi_{1}}\otimes\ket{\psi_{2}}$ where the subsystem $A$ belongs to state $\ket{\psi_{1}}$, while $B$ belongs to $\ket{\psi_{2}}$. Single-qubit measurements outside of the $A$ and $B$ subsystems cannot change the mutual information between these subsystems, since they belong to a separable state, so the entangling power $\Delta I_{AB} = 0$.    More generally in the separable phase, the entangling power, which decays at least as fast as the probability that two randomly chosen spins belong to an entangled cluster in the separable phase, will vanish as $\overline{\Delta I_{AB}} \lesssim 1/N$, as $N\rightarrow\infty$. 

\begin{figure}
\includegraphics[trim = 0 0 0 0, clip = true, width=0.17\textwidth, angle = 0.]{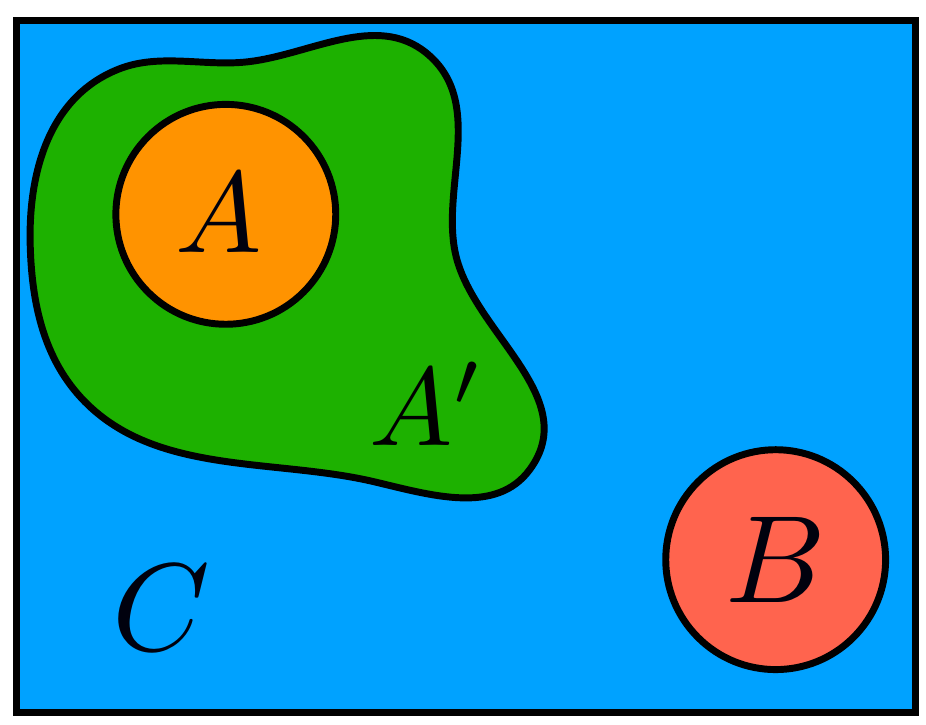}
\includegraphics[trim = 0 0 0 0, clip = true, width=0.17\textwidth, angle = 0.]{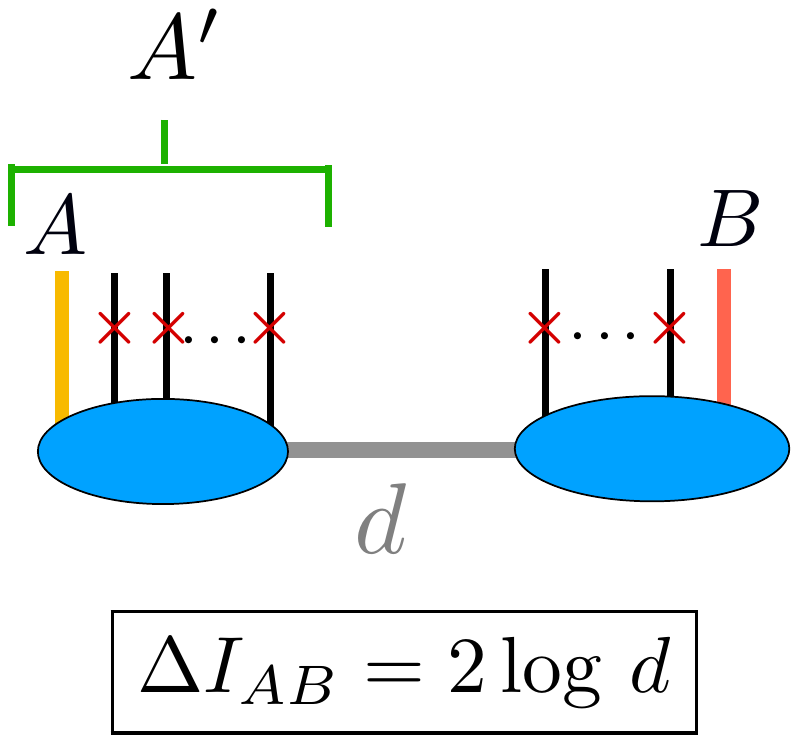}
\caption{{\bf The Entangling Power:} The separability of the steady-state is probed by a measuring the change in the mutual information between two finite subsystems $A$ and $B$, after measuring each of the spins outside of these two subsystems in an arbitrary basis.  In a random tensor network, the entangling power reveals the minimal extent to which two subsystems are entangled. }
  \label{fig:RTN_Example}
\end{figure}

In the fully-entangled phase, however, we may argue that the entangling power is generally be non-zero by considering specific examples.  First, we consider the behavior of the entangling power in a random tensor-network state for $N\gg 1$ spins.  For such a state, the mutual information between two finite subsystems $A$ and $B$ will typically be small $I_{AB} \approx 0$.  We now consider an arbitrary subsystem $A'$ which include the $A$ subsystem $(A\subseteq A')$ and is disjoint with $B$, as illustrated in Fig. \ref{fig:RTN_Example}.  Subject to these conditions, we choose $A'$ such that its von Neumann entanglement entropy is \emph{minimized} and is given by $S_{A'} = \log d$.   The von Neumann entanglement entropy may be obtained by considering a ``minimal cut" through the bonds of the tensor network state when $d_{A}, d_{B},d\gg 1$, where $d_{A,B}$ are the Hilbert space dimensions of $A$ and $B$ \footnote{The entanglement $S_{A} = \min_{\gamma}[\sum_{i\in\gamma}\log d_{i}]$ where $\gamma$ is a path that partitions the degrees of freedom in $A$ from the rest of the system, and $d_{i}$ is the dimension of bond $i$ crossed by the path}; this is an upper bound to the entanglement which is saturated when the bond dimensions of the tensor network are sufficiently large.   This implies that the state may be written as the product of two tensors with bond dimension $d$, $\ket{\Psi} = \sum_{s=1}^{d}V^{\boldsymbol{\sigma}_{A'}}_{s}W^{\boldsymbol{\mu}_{\bar{A'}}}_{s}\ket{\boldsymbol{\sigma}_{A'}, \boldsymbol{\mu}_{\bar{A'}}}$ as shown in Fig. \ref{fig:RTN_Example}.  We now perform projective measurements of all of the degrees of freedom outside of the $A$ and $B$ subsystems.  From the minimal cut prescription and the particular choice of the $A'$ subsystem, the mutual information after performing these measurements  is $2 \log d$.  Equivalently, the change in the mutual information 
\begin{align}\label{eq:IAB_change}
\Delta I_{AB} = 2S_{A'} = 2\log d.
\end{align}
Therefore, the entangling power reveals the bottleneck in the entanglement between the two subsystems, i.e. the minimal extent to which these subsystems are truly entangled ``with each other".  We note that the minimal cut prescription in this example is independent of the measurement basis, and the measurement outcomes.  %This example also shows that the % When the bond dimensions of the tensor network state not large, Eq. (\ref{eq:IAB_change}) provides an upper bound to the change in the mutual information.
Apart from random tensor networks, we have checked that other fully-entangled states also exhibit this behavior. In code states of the 5-qubit code \cite{Gottesman1996Class} for example, any pair of spins has zero mutual information. Taking $A$ and $B$ to be two random spins, we have checked that performing measurements of the remaining spins in a random basis  generically boosts their mutual information.

\begin{figure}
$\begin{array}{c}
\includegraphics[trim = 0 0 0 0, clip = true, width=0.31\textwidth, angle = 0.]{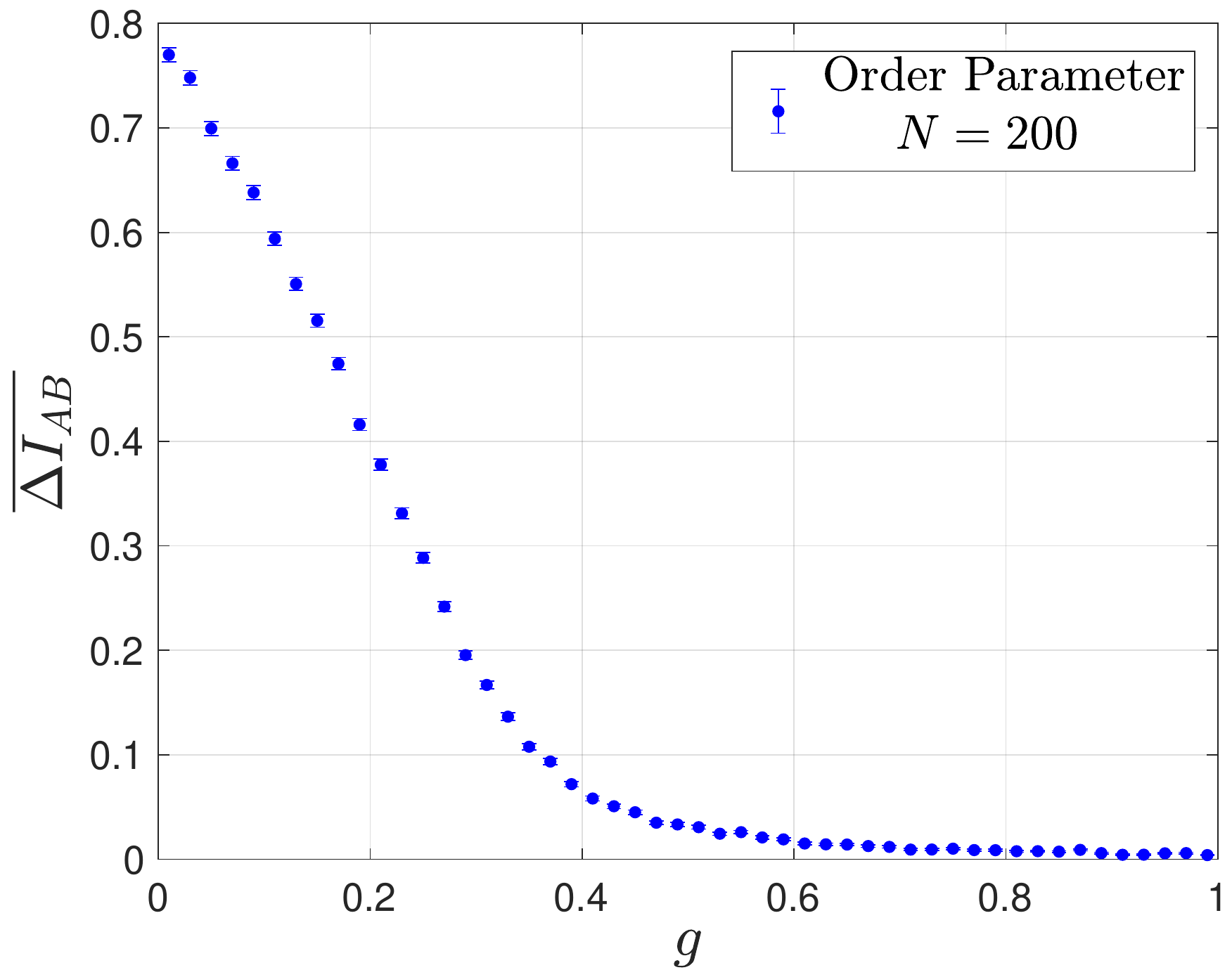}\\
\text{(a)}\\
\includegraphics[trim = 0 0 0 0, clip = true, width=0.31\textwidth, angle = 0.]{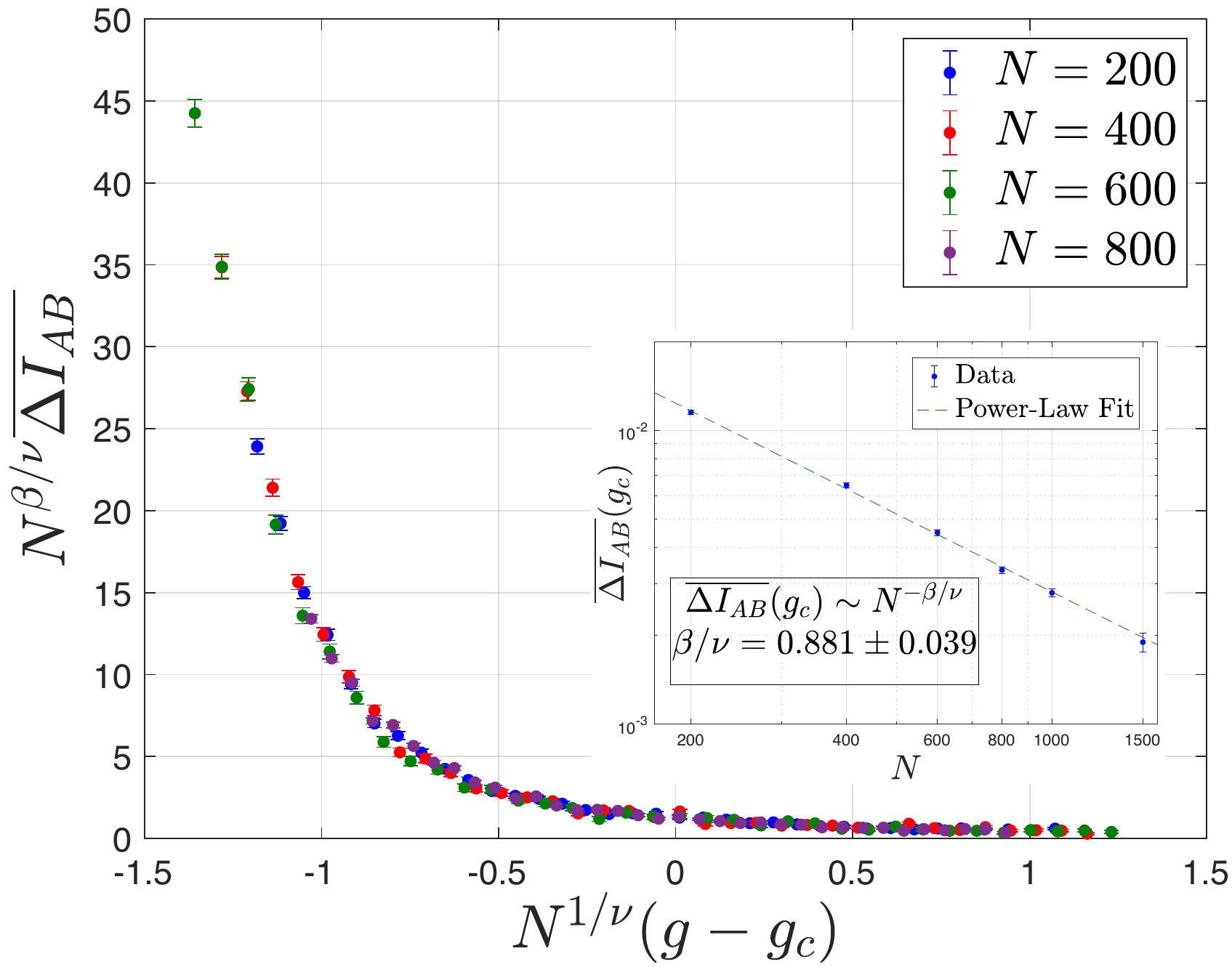}\\
\text{(b)}
\end{array}$
\caption{{\bf Numerical Study of the Order Parameter:}  In our toy model with Clifford dynamics, the proposed order parameter appears to distinguish between the fully entangled and separable phases as shown in (a).  In (b), we show a scaling collapse of the order parameter for the indicated system sizes from Eq. (\ref{eq:scaling_form_order_param}), which leads to the critical exponents quoted in the main text.}
  \label{fig:Order_Parameter}
\end{figure}

\emph{Critical Behavior of the Order Parameter}: We now study the behavior of the entangling power in the Clifford dynamics discussed previously.  We choose the $A$ and $B$ subsystems to each be a randomly chosen spin.  We measure the entangling power by computing the change in the mutual information between these two spins after measuring the remaining spins in the Pauli $X$ basis, and averaging over
$O(10^{5})$ iterations.  The behavior of the order parameter as a function of $g$ is shown in Fig. \ref{fig:Order_Parameter}a for $N = 200$ spins.   

We further propose the finite-size scaling form for the entangling power near criticality
\begin{align}\label{eq:scaling_form_order_param}
\overline{\Delta I_{AB}}(g, N) = N^{-\beta/\nu}h(N^{1/\nu}|g - g_{c}|)
\end{align}
From fitting the value of $\overline{\Delta I_{AB}}$ at criticality ($g = g_{c}$) as a function of system size, we find that $\beta/\nu = 0.881 \pm 0.039$ as shown in the inset in Fig. \ref{fig:Order_Parameter}b.  A finite-size scaling collapse for $1/\nu = 0.18$ is also shown in Fig. \ref{fig:Order_Parameter}b for various system sizes.  %We note that if the entangling power were strictly given by the probability that the two spins $A$ and $B$ belong to the same, entangled cluster, then we would predict that $\beta/\nu = 1$.  

{\bf\emph{Computational ``Hardness" of the Entangled and Separable Phases:}}  In one spatial dimension, the measurement-driven transition between asymptotically volume- or area-law scaling of the entanglement entropy naturally corresponds to a transition in the computational difficulty in simulating these dynamics; the resources required to store the wavefunction of a many-body system scale with the entanglement of a bipartitioning, making simulation of the dynamics of a generic, volume-law entangled state unfeasible in a large system \cite{Skinner2019MPTDE,napp2019efficient}.  Moreover, sampling the output probability distribution of a many-body wavefunction in a fixed basis is classically unfeasible at sufficiently long times if the state is evolving under random unitary gates \cite{Harrow_Supremacy}.  For all-to-all dynamics involving projective measurements, and generic (e.g. Haar-random) few-body unitary gates, we are not aware of a classically-efficient way to simulate the dynamics as the rate of projective measurements is increased, even if the system undergoes a separability transition.  Nevertheless, for the IQP dynamics we have described, in which the applied unitary gates are random, non-Clifford unitary gates that mutually commute, there is a sampling task whose difficulty coincides with the separability phase transition. 

 While the wavefunction (\ref{eq:graph_evolution}) for a state generated by an IQP circuit is clearly classically efficient to track and store, determining the probability distribution for this state over product states in the Pauli $X$ basis can still be classically hard \cite{IQP_Hard}.  In the separable phase, this task is substantially simplified by the fact that ($i$) this distribution is given by the product of the distributions for each of the finite, separable components of the state and ($ii$) the structure of the wavefunction permits an efficient calculation of this distribution for each separable component.  An estimate of the ``hardness" of this calculation is as follows: the size of the largest fully-entangled cluster in the separable phase of the system is $O(s(g)\log N)$ due to the exponential decay of the cluster size distribution in Eq. (\ref{eq:power_law}), where $s(g) \sim |g-g_{c}|^{-2}$ near criticality in a thermodynamically large system \footnote{We observe that the largest component with size $N m(g)$ satisfies the equation $N\sum_{k\ge Nm(g)}n_{k}(g) = 1$, where $n_{k}(g)$ is the cluster size distribution which decays as $n_{k}(g) \sim k^{-5/2}e^{-k/s(g)}$  when $g>g_{c}$ so that $N\sum_{k\ge Nm(g)}n_{k}(g) \sim Ne^{-Nm/s(g)} (Nm)^{-3/2}$.  In the $N\rightarrow\infty$ limit, this requires that $N\,m\sim s(g)\log N$.}.  
  
  A naive assumption for the difficulty of calculating the probability for each separable component, would be that it would take a time exponentially long in the size of each component.   This estimate is substantially improved by observing that in the separable phase, a unitary operator will typically be applied between two spins that belong to clusters that are disentangled from each other.  As a result, the graph $\boldsymbol{G}(t)$ corresponding to the separable phase will consist of trees, i.e. graphs for which there are no loops, and for which there is a unique path connecting any two nodes, so that the total number of paths connecting all pairs of points scales {quadratically} in the size of the tree.  Probabilities such as $|\braket{\rightarrow\cdots\rightarrow|\Psi(t)}|^{2} = |\sum_{\boldsymbol{s}}\Psi(\boldsymbol{s};t)|^{2}$ require calculating sums over paths on the graph $\boldsymbol{G}(t)$ with complex weights.  With this, the typical size of a large entangled cluster in the separable phase, and assuming that there are $O(N/s(g)\log N)$ such clusters, we obtain the estimate that only $O(s(g) N\log N)$ operations would be required to calculate the probability distribution for the steady-state in the Pauli-$X$ basis.  In a finite system, the critical point is broadened into a scaling window $|g - g_{c}| \sim N^{-1/3}$ \cite{luczak1990component, bollobas1984evolution}.  This implies that $s(g)$ diverges as $s(g) \sim N^{2/3}$ near criticality, and leads to an esetimate $O(N^{5/3}\log N)$ operations required to determine the probability of being in a particular state in the Pauli-$X$ basis.  In contrast, in the fully-entangled phase, the largest fully-entangled cluster of spins covers a finite fraction of the system, and determining these probabilities require calculating the ``partition function" for an Ising model with complex weights on a random graph with $O(N)$ nodes. Since the degree of each node in the graph is finite, the graph will locally appear to be tree-like.  Nevertheless, the graph corresponding to this cluster will consist of long loops of length $\ell \gg \log N$.  We are unaware of an efficient way that this can be performed \cite{IQP_Hard, IQP_Hard2}.   
 
 % Assuming that calculating the probability for each of the $O(N/s(g)\log N)$ components takes an exponentially long time in the size of each component, we obtain that the number of operations required to determine the probability of being in a particular state in the Pauli $X$ basis, given knowledge of the state (\ref{eq:graph_evolution}) is at worst $O(N^{1+c\,s(g)}/(s(g)\log N))$ where $c$ is a finite constant, independent of $N$ and $g$.  
 
%; in the fully-entangled phase, however, t

We conclude by making these ideas more precise through a concrete experiment. In the steady-state, we apply a single-qubit unitary operation $U(t)$ which restores an Ising symmetry to the state, so that  $\ket{\Phi(t)} \equiv U(t)\ket{\Psi(t)}$ satisfies $\braket{\Phi(t)|\prod_{m}X_{m}|\Phi(t)} = 1$ at all times.  Concretely, $U(t) \equiv \sum_{\boldsymbol{s}}\exp[-i\boldsymbol{s}\cdot \boldsymbol{z}(t)] \ket{\boldsymbol{s}}\bra{\boldsymbol{s}}$, where the vector $\boldsymbol{z}$ has components $z_{m}(t) \equiv w_{m}(t) + (1/2)\sum_{n}\theta_{mn}(t)$.  We claim that the probability 
\begin{align}
P(t) \equiv |\braket{\rightarrow\cdots\rightarrow\, |\, \Phi(t)}|^{2}
\end{align}
can be efficiently calculated in the separable phase as follows.  %First, we note that the separable phase, a unitary operator will typically be applied between two spins that belong to clusters that are disentangled from each other.  As a result, the graph $\boldsymbol{G}(t)$ corresponding to the separable phase will consist of trees, i.e. graphs for which there are no loops. 
 From the form of the state $\ket{\Phi(t) }$ 
\begin{align}
\ket{\Phi(t) } \equiv \frac{1}{\sqrt{D}}\sum_{\boldsymbol{\sigma}}\prod_{n,m}\exp\left[\frac{i}{8}\sigma_{n}\theta_{nm}(t)\sigma_{m}\right]\ket{\sigma_{1},\ldots,\sigma_{N}}\nonumber
\end{align}
where $\sigma_{n} = \pm 1$ denotes the state of spin $n$ in the Pauli $Z$ basis, the return probability $P(t)$ on a \emph{tree} graph is exactly 
\begin{align}
P(t) = \prod_{n<m}\cos^{2}[\theta_{nm}(t)/4]
\end{align}
This quantity should agree with the true return probability, deep in the separable phase, and deviate from this in the fully-entangled phase, which could provide an additional signature of this transition.

\emph{Acknowledgments:} SV thanks Ruihua Fan, Ashvin Vishwanath, Yi-Zhuang You, and Adam Nahum for useful discussions and for collaboration on previous related work. SV is supported by the Harvard Society of Fellows.

\bibliography{references}

\onecolumngrid
\newpage
\appendix
\section{Deriving the Dynamical Rules for the Evolving State}%\label{app:dynamical_rules}
We consider the dynamics of the initial state $\ket{\rightarrow\cdots\rightarrow}$ under randomly-applied one- and two-qubit  gates that mutually commute with each other, and that generate no entanglement in the Pauli-$Z$ basis.  Let $U_{\mathrm{IQP}}$ be the unitary circuit generated by the continued application of these gates.  Since Pauli-$Z$ measurements commute with this unitary operator, the probability of measuring a single spin up or down in the $Z$ basis is time-independent, and identical to that of the initial state (for the projection operator $P^{(\pm)}_{i} \equiv (1\pm Z_{i})/2$,  observe that $\braket{\phi | U_{\mathrm{IQP}}^{\dagger}P^{(\pm)}_{i}\,U_{\mathrm{IQP}} |\phi} = \braket{\phi | P^{(\pm)}_{i}|\phi}$, for any state $\ket{\phi}$).  

The action of $U_{\mathrm{IQP}}$ on a product state in the Pauli-$X$ basis is, up to an overall phase,  
\begin{align}\label{eq:IQP_state}
\ket{\psi}\equiv U_{\mathrm{IQP}}\ket{\rightarrow\cdots\rightarrow} = 2^{-N/2}\sum_{\boldsymbol{s}}\exp\left[{\frac{i}{2}\boldsymbol{s}^{T}\boldsymbol{\theta}\boldsymbol{s} + i\boldsymbol{v}\cdot\boldsymbol{s}}\right]\ket{s_{1},\ldots,s_{N}}
\end{align}
where $\boldsymbol{s} = (s_{1},\ldots,s_{N})$ is a representation of the state of the $N$ spins in the $Z$ basis, where $s_{1} = 0$ or $1$ if the spin is up or down, respectively.  Here, $\boldsymbol{\theta}$ is a symmetric $N\times N$ matrix and $\boldsymbol{v}$ is an $N$-component vector. 
Now, say we perform a Pauli-$Z$ measurement on the first spin, followed by a rotation of the spin into the $\ket{\rightarrow}$ state so that it can continue to entangle with the other spins under the unitary dynamics.  The state of the system $\ket{\psi_{s}}$ after (i) the  measurement which yields the result that the first spin is in the state $s\in\{0,1\}$,  and (ii) the rotation into the Pauli $X$ basis, is given by 
\begin{align}
\ket{\psi_{s}} = \exp\left[\frac{i\pi}{4}(2s-1)Y_{1}\right] \frac{P_{s}\ket{\psi}}{\sqrt{\braket{\psi|P_{s}|\psi}}} = \sqrt{2}\exp\left[\frac{i\pi}{4}(2s-1)Y_{1}\right]P_{s}\ket{\psi}
\end{align}
where we have defined the projection operator $P_{s} = [1-(2s-1)Z_{1}]/2$.  We note that $\ket{\psi_{s}}$ may be written exactly as in Eq. (\ref{eq:IQP_state}), with a re-definition of $\boldsymbol{\theta}$ and $\boldsymbol{v}$ as 
\begin{align}
\theta_{ij} \longrightarrow \theta'_{ij}\equiv \left\{\begin{array}{cc}
\theta_{ij} & i\ne1\,\text{and}\,j\ne1\\
0 & i=1\,\text{or}\,j=1
\end{array}\right.
\hspace{.25in}
v_{i}\longrightarrow v'_{i} \equiv \left\{\begin{array}{cc} 
v_{i} + s\,\theta_{1i} & i \ne 1\\
0 & i=1
\end{array}\right.
\end{align}
Therefore, the resulting state can be described by another IQP circuit consisting of single- and two-qubit gates, that acts on the same initial state.  We observe that the measurement outcome only affects the single-qubit gates in this new IQP circuit; as a result, the outcome of the measurement doesn't change the entanglement properties of the resulting state.

For a generic IQP state of the form given in Eq. (\ref{eq:IQP_state}), we may calculate the purity for a bipartitioning that divides the system into a subsystem $A$ and its complement $\bar{A}$.  We may write $\boldsymbol{\theta}$ in block form as
\begin{align}
\boldsymbol{\theta} = \left(\begin{array}{cc}
\boldsymbol{\theta}_{A} & \boldsymbol{\varphi}\\ \\
\boldsymbol{\varphi}^{T} & \boldsymbol{\theta}_{\bar{A}}
\end{array}\right)
\end{align}
where $\boldsymbol{\theta}_{A}$ and $ \boldsymbol{\theta}_{\bar{A}}$ each act exclusively within the $A$ and $\bar{A}$ subsystems, respectively, while $\boldsymbol{\varphi}$ is an $N_{A}\times N_{\bar{A}}$ matrix that acts across the subsystems.  Furthermore, let $\boldsymbol{r}$, $\boldsymbol{r'}$ denote a representation of the state of the $N_{\bar{A}}$ spins in the $\bar{A}$ subsystem in the $Z$ basis, with $r_{i} = 0$ or $1$ if the corresponding spin points up or down, respectively. With this decomposition, we may write the purity of the density matrix $\rho_{A} \equiv \Tr_{\bar{A}}\ket{\psi}\bra{\psi}$ as
\begin{align}
\Tr(\rho_{A}^{2}) = \frac{D_{A}}{D^{2}}\sum_{\boldsymbol{r},\boldsymbol{r}'}\,\prod_{i\in A}\left[1 + \cos\Big(\sum_{j\in \bar{A}}\varphi_{ij}[{r}_{j} - r'_{j}]  \Big)\right]
\end{align}
From the above expression, if the matrix $\varphi$ is not the zero matrix (mod $2\pi$), then $\Tr(\rho_{A}^{2}) < 1$ and the region $A$ is entangled with the rest of the system.

\subsection{Clifford Dynamics}
We now derive these dynamical rules for the special case of Clifford dynamics.  First, in the absence of any measurements, we observe that the wavefunction may be uniquely specified by  $N$ evolving {stabilizers}, $\{\mO_{1}(t),\ldots,\mO_{N}(t)\}$ which are defined as mutually commuting Pauli operators which each satisfy 
\begin{align}
\mathcal{O}_{m}(t)\ket{\Psi(t)} = \ket{\Psi(t)} \hspace{.2in} m \in \{1,\ldots,N\}
\end{align} 
and with $\mathcal{O}_{m}(0) = X_{m}$.  

\begin{figure*}
$\begin{array}{ccc}
\includegraphics[trim = 0 0 0 0, clip = true, width=0.3\textwidth, angle = 0.]{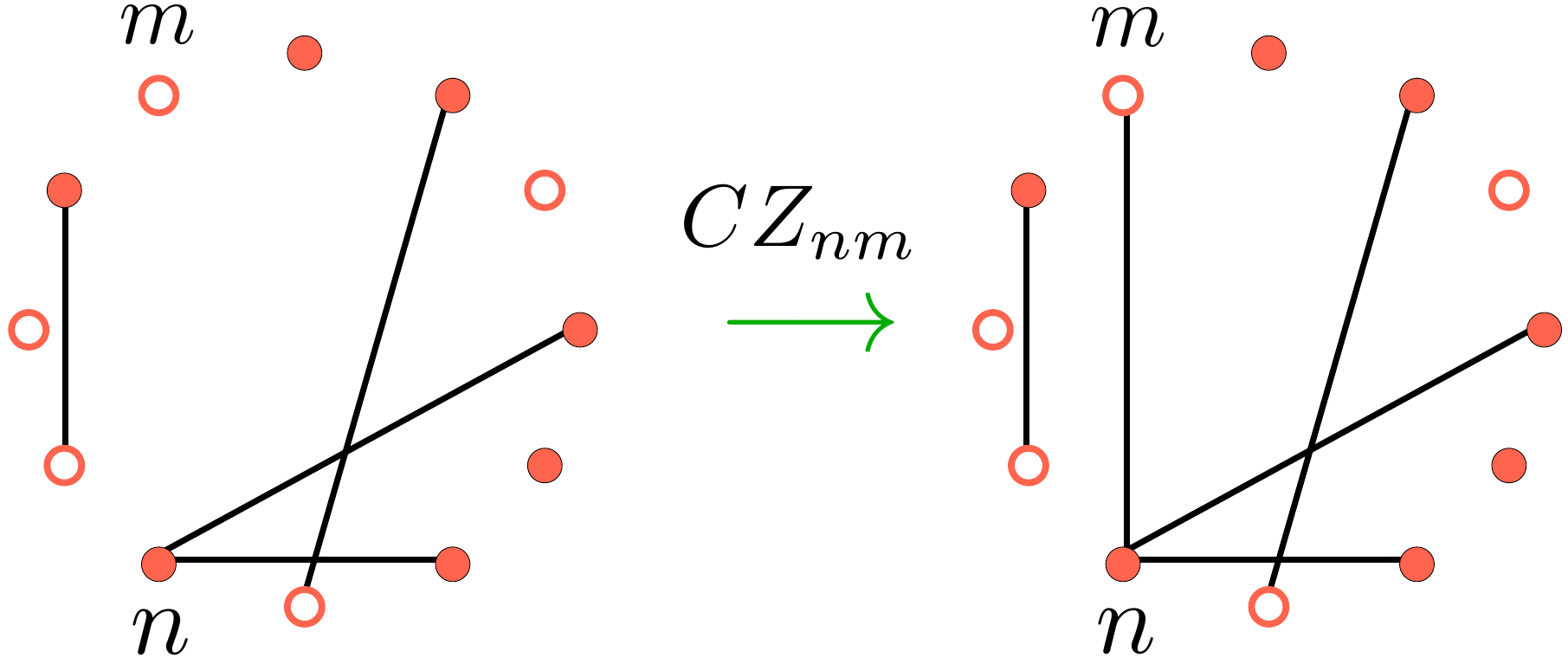}\text{\hspace{.25in}} & &
\includegraphics[trim = 0 0 0 0, clip = true, width=0.32\textwidth, angle = 0.]{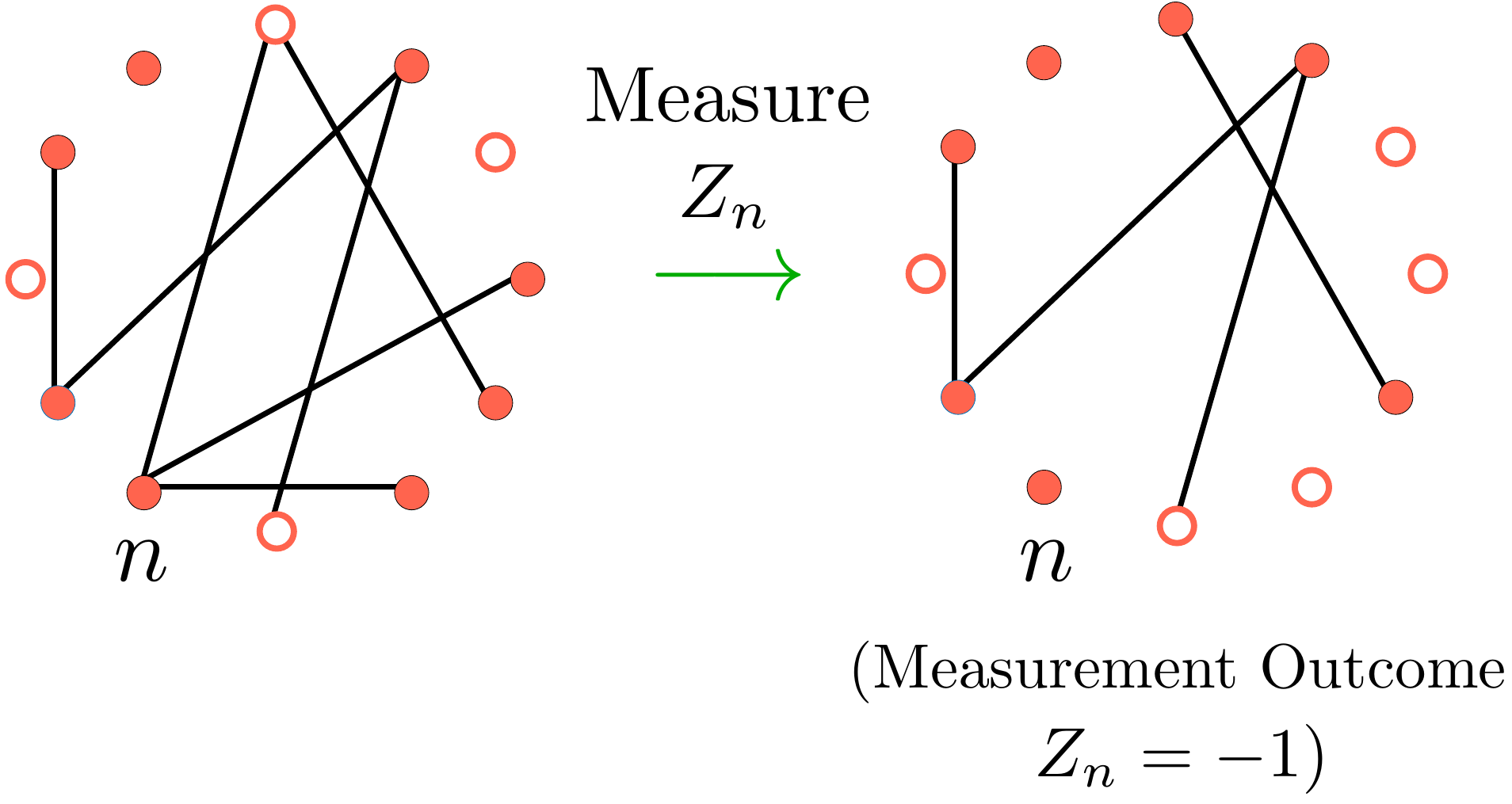}\\
\text{(a)} & & \text{(b)}
\end{array}$
\caption{{\bf Dynamics of the Evolving State:} Evolution of the wavefunction under (a) $CZ$ operations and (b) single-site measurements in the Pauli $Z$ basis, followed by a rotation to the $X$ basis.  The wavefunction is specified by a simple graph with $N$ nodes that can be empty or filled; the graph connectivity and the node colorings are encoded in an adjacency matrix $G$ and in a binary vector $v$, respectively.  Applying $CZ_{nm}$ corresponds to adding a bond between nodes $n$ and $m$ if none is already present, or removing the bond if a bond is present, while a measurement at site $n$ removes all bonds from node $n$.  If the measurement of spin $n$ yields $Z_{n} = -1$, as has been assumed in (b), then the colorings of the nodes that were previously attached to $n$ are flipped. }
  \label{fig:Dynamics}
\end{figure*}

We now consider evolving the initial state with the unitary operator
\begin{align}
U_{\boldsymbol{G}} \equiv \prod_{n<m} \left(CZ_{nm}\right)^{G_{nm}}
\end{align}
where the binary matrix $\boldsymbol{G}$ has entries $G_{nm} = 1$ if a control-$Z$ is to be applied between spins $n$ and $m$, and is zero otherwise.   Since $CZ_{nm}X_{n}CZ_{nm} = X_{n}Z_{m}$, the Heisenberg evolution of the stabilizers is given by
\begin{align}
U_{\boldsymbol{G}}X_{n} U_{\boldsymbol{G}}^{\dagger} = X_{n}\prod_{m}(Z_{m})^{G_{nm}}
\end{align}
The control-$Z$ operators mutually commute and each square to the identity.  Therefore, $U_{\boldsymbol{G}}U_{\boldsymbol{G'}} = U_{\boldsymbol{G}\oplus \boldsymbol{G'}}$, where $\boldsymbol{G}\oplus \boldsymbol{G'}$ denotes the {binary} addition of the two matrices; the evolution of the initial state under control-$Z$ gates is then described by the dynamics of a {simple} graph with $N$ nodes. A control-$Z$ operator corresponds to adding or removing a bond between the corresponding nodes in the graph, as shown in Fig. \ref{fig:Dynamics}. The wavefunction defined by the stabilizers $X_{n}(t) = U_{\boldsymbol{G}(t)}X_{n}U^{\dagger}_{\boldsymbol{G}(t)}$ is given by Eq. (\ref{eq:graph_evolution}) in the Pauli $Z$ basis, with $\boldsymbol{v}=0$.

We now consider how the stabilizers evolve after performing a measurement. At a time $t$, we measure the first spin in the $Z$ basis, and obtain a measurement outcome $(-1)^{z_{1}}$, with  $z_{1} \in \{0, 1\}$.  This measurement manifestly commutes with all stabilizers $X_{n}(t) = U_{\boldsymbol{G}(t)}X_{n}U^{\dagger}_{\boldsymbol{G}(t)}$ except for ${X}_{1}(t)$.  After performing the measurement, the new state is described by the following stabilizers: ${X}_{1}(t)$ is replaced by the stabilizer $(-1)^{z_{1}}Z_{1}$, while the remaining stabilizers are modified by replacing any appearance of the operator $Z_{1}$ in each stabilizer, with the measurement outcome $(-1)^{z_{1}}$.  %This may be succinctly written as ${X}_{n}(t)  \rightarrow (-1)^{z_{1}G_{n1}(t)} (Z_{1})^{G_{n1}}X_{n}(t)$ when $n\ne 1$.    
We then rotate the measured spin into the Pauli $X$ basis, so that $X_{1}$ becomes a stabilizer for the new state.  To summarize, the stabilizers for the new state may be simply written as 
\begin{align}\label{eq:new_stab}
X_{m}(t) = (-1)^{v_{m}(t)}U_{\boldsymbol{G'}(t)}X_{m}U_{\boldsymbol{G'}(t)}^{\dagger}
\end{align}
with $m \in \{1,\ldots, N\}$.  Here, $v_{m}(t) \equiv z_{1}G_{m1}(t)$, and the adjacency matrix $\boldsymbol{G'}(t)$ is obtained from $\boldsymbol{G}(t)$ by deleting all of the bonds connected to the first node.  It is straightforward to verify that these operators stabilize the wavefunction in Eq. (\ref{eq:graph_evolution}), with adjacency matrix $\boldsymbol{G'}(t)$.

In the presence of both ($i$) control-$Z$ operators and ($ii$) measurements in the Pauli-$Z$ basis, followed by a rotation to the $X$ basis, it is clear that the stabilizers for the evolving wavefunction will always take the general form in Eq. (\ref{eq:new_stab}).  The wavefunction
\begin{align}
|\Psi(t)\rangle \equiv \prod_{m}\left[\frac{1 + X_{m}(t)}{2}\right]|\uparrow\cdots\uparrow\rangle
\end{align}
then has the matrix elements given in Eq. (\ref{eq:graph_evolution}) in the main text.
Therefore, the state of the system is specified by an evolving graph with ($i$) adjacency matrix $\boldsymbol{G}$, and ($ii$) nodes that are filled or empty, depending on whether $v_{m} = 0$ or $1$, respectively.  Applying $CZ_{nm}$ to the state simply removes the bond between nodes $n$ and $m$ if one is already present, and adds a bond otherwise.  A measurement of spin $k$ in the Pauli $Z$ basis will always commute with all stabilizers except $X_{k}(t)$, so that the measurement, followed by a subsequent rotation to the Pauli $X$ basis, corresponds to removing all bonds connected to node $k$ in the graph.  If the measurement yielded the result that the spin is pointing down, then the filling of all nodes that were previously connected to node $k$ are changed.  

We note that for the stabilizers in Eq. (\ref{eq:new_stab}), the overall phase factor may be removed by applying a unitary transformation $W = \prod_{m}(Z_{m})^{v_{m}(t)}$, which acts independently on each spin, so that the spectrum of any reduced density matrix for the state is identical to that of a state with $\boldsymbol{v} = 0$.  Therefore, the entanglement properties of the state only depend on the graph's adjacency matrix.

\section{Rate Equation for the Graphical Evolution}%\label{app:rate_eq}
Substituting the expressions for the rates $\gamma_{k'\rightarrow k}$ into Eq. (\ref{eq:stab_evolution}) yields the rate equation
\begin{align}
\frac{ds_{k}}{dt} = \,&2\,\Gamma_{u}\left(s_{k-1}-s_{k}\right) + \Gamma_{m} k\left(s_{k+1}-s_{k}\right)\nonumber\\
&+\Gamma_{m}\delta_{k,1}
\end{align}
Introducing the generating function $F(z,t) \equiv \sum_{k\ge 1}z^{k}s_{k}(t)$, we observe that the generating function satisfies the equation
\begin{align}
\frac{\partial F}{\partial \tau} = \left[z-1-\frac{g}{z}\right]F(z,\tau)  + g(1-z) \frac{\partial F}{\partial{z}}+ zg
\end{align}
where $\tau \equiv 2\Gamma_{u}t$ and $g \equiv \Gamma_{m}/(2\Gamma_{u})$.  Given the initial condition $F(z,0) = z$, we obtain the solution
\begin{align}
F(z,\tau) = \frac{zg}{1-z}\left\{1 - \left[1 - \frac{1-z}{g}e^{-g\tau}\right]e^{-(1-z)\left(1-e^{-g\tau}\right)/g}\right\}\nonumber
\end{align}
The steady-state, $f(z) \equiv \lim_{\tau\rightarrow\infty}F(z,\tau)$ is then given by
\begin{align}\label{eq:steady_state_gen_function}
f(z) = \frac{zg}{1-z}\left[1 - e^{-(1-z)/g} \right]
\end{align}
Expanding the generating function in the steady-state about $z=0$, we find that
\begin{align}
s_{k}^{(\infty)} = \lim_{\tau\rightarrow\infty} s_{k}(\tau) = g\left[ 1 - \frac{\Gamma\left(k, 1/g\right)}{(k-1)!}\right]
\end{align}
where
\begin{align}
\Gamma(k,a) = \int_{a}^{\infty}dx\,x^{k-1}e^{-x}
\end{align}
is the incomplete gamma function.  The asymptotic form of this distribution is 
\begin{align}
s_{k}^{(\infty)} = g\frac{(e/g)^{k}e^{-1/g}}{k^{k}\,\sqrt{2\pi k}}\left( 1 + O(k^{-1})\right)
\end{align}
when $k \gg 1$.

\section{Size of the Largest Entangled Cluster}%\label{app:largest_ent_cluster}
We determine analytic expressions for the size of the largest entangled cluster of spins.  We note that the equations that we derive below were rigorously obtained originally in Ref. \cite{Strogatz1}. We consider a state, as described in the text, where the nodes of the graph $\boldsymbol{G}$ have degrees drawn from the distribution $s_{k}^{(\infty)}$.  The degree distribution of a random node in this graph is therefore $s_{k+1}^{(\infty)}$, with $k \ge 0$. The probability $p$ that a randomly chosen node belongs to a cluster of finite size is equal to the probability that all of the node's \emph{neighbors} also belong to finite-sized clusters.  Therefore, 
\begin{align}\label{eq:MR1}
p = \sum_{k\ge 0} s_{k+1}^{(\infty)} q^{k}
\end{align}
Here, $q$ is the probability that a node (say $A$) that is reached {by following a random bond}, belongs to a finite-sized cluster.  We now determine $q$ as follows.  First, recall that $A$ has the normalized degree distribution $P_{k} = 2g k s_{k+1}^{(\infty)}$, for $k \ge 0$.  We now follow the bonds from $A$ to determine the the probability that $A$'s other neighbors also belong to finite clusters. If $A$ has degree $k$, then there are $k-1$ bonds other than the bond that we followed to reach $A$ that have not been traversed; the probability that the nodes attached to each of these $k-1$ bonds all belong to a finite cluster is equal to the probability that $A$ does, as well.  Summing over all possible bond configurations for $A$ leads to the condition that%\footnote{In this analysis, we assume that a finite cluster in the graph will have a tree-like structure}  
\begin{align}\label{eq:MR2}
q = \sum_{k\ge 0} P_{k+1}q^{k}
\end{align}

These equations may be solved to obtain the mass of the infinite cluster 
\begin{align}
m = 1-p.
\end{align}
From the form of the fixed-point stabilizer distribution, $q$ is given by the solution to the equation
\begin{align}\label{eq:q_eqn}
q(1-q)^{2} + 2g\left[e^{-(1-q)/g}(1 - q + g) - g\right] = 0
\end{align}
For various values of $g$, we solve Eq. (\ref{eq:q_eqn}) numerically, and substitute this into the expression for the cluster mass $m(g) = 1- p$, which is given by
\begin{align} 
m(g) = 1 - \frac{g}{1-q}\left(1 - e^{-(1-q)/g}\right)
\end{align}
in order to obtain the result plotted in Fig. \ref{fig:Cluster_Mass}.  It is possible to analytically derive the form of the cluster mass near criticality from these equations
\begin{align}
m(g) = 2(g_{c} - g) + O(g_{c} - g)^{2}
\end{align}
as $g \rightarrow g_{c}^{-}$.

%\section{Condition for the validity of Eq. (\ref{eq:power_law})}
%In Ref. \cite{Strogatz1}, a condition is derived for the validity of the power-law decay of the cluster size distribution $n_{k} \sim k^{-5/2}$ for a random graph with an arbitrary node degree distribution.  This condition is given in terms of the generating function $G(z)$ for the probability distribution $p_{k} = k n_{k}$ that a random node belongs to a cluster of size $k$, which is defined as
%\begin{align} 
%G(z) = \sum_{k\ge 0} z^{k} p_{k}
%\end{align}
%The condition derived in Ref. \cite{Strogatz1} is simply that $G'''(1)/G'(1)$ is non-zero.  Using the fact that $G(z) = f'(z) - z^{-1}f(z)$, where $f(z)$ is given in Eq. (\ref{eq:steady_state_gen_function}), we observe that
%$G'''(1)/G'(1) = (4 + 15 g)/(10 g^3)$, which non-zero for all $g$.

\section{Bounding the Entanglement Entropy}
We derive a simple bound on the entanglement entropy in the dynamics described in the main text.  Assuming that each of the $k_{AB}$ bonds ends at an arbitrary node in the $B$ sub-system, we calculate that the probability that $s$ distinct sites in the $B$ sub-system are reached by these bonds.  First, there are $\left(\begin{array}{c} N-m\\ s \end{array} \right)$ choices of $s$ sites in the $B$ sub-system.  The number of ways that the $k_{AB}$ bonds can be configured to cover a given choice of $s$ sites is precisely 
\begin{align}\label{eq:sum_sites_B}
&\sum_{n_{1} + \cdots + n_{s} = k_{AB};\,\, n_{i}\ge 1}\frac{(k_{AB})!}{n_{1}!\cdots n_{s}!} = s! \left\{ \begin{array}{c} k_{AB}\\s\end{array}\right\}
\end{align}
where $\Big\{\begin{array}{c}a\\b\end{array}\Big\}$ denotes the Stirling number of the second kind, which counts the number of partitions of $a$ elements into $b$ non-empty subsets.  This is because, for a fixed configuration of $s$ sites in the $B$ subsystem, let $\{n_{1}, \ldots, n_{s}\}$ denote a configuration of the $k_{AB}$ bonds, with $n_{i}$ of the bonds reaching site $i\in\{1,\ldots,s\}$, respectively.  For a fixed set of these integers, there are
\begin{align}
\prod_{i=1}^{s}\left(\begin{array}{c} k_{AB} - \displaystyle\sum_{j < i}n_{j}\\ n_{i}\end{array}\right) = \frac{(k_{AB})!}{n_{1}!\cdots n_{s}!}
\end{align}
different ways to choose the $k_{AB}$ bonds.  Summing over all positive integers $\{n_{1}, \ldots, n_{s}\}$, subject to the constraint that $n_{1} + \cdots + n_{s} = k_{AB}$ gives Eq. (\ref{eq:sum_sites_B}).  From this, we conclude that the probability that $s$ sites in the $B$ sub-system are reached is precisely given by
\begin{align}
P(s) = \frac{ s! \left\{ \begin{array}{c} k_{AB}\\s\end{array}\right\} \left(\begin{array}{c}N-m\\s\end{array}\right)}{\displaystyle\sum_{s=0}^{N-m} s! \left\{ \begin{array}{c} k_{AB}\\s\end{array}\right\} \left(\begin{array}{c}N-m\\s\end{array}\right)}
\end{align}
We may evaluate the denominator explicitly by using the following representation of the Stirling number of the second kind
\begin{align}\label{eq:StirlingS2_identity}
m! \left\{ \begin{array}{c} n\\m\end{array}\right\} = \left(\frac{d}{dx}\right)^{n} \left(e^{x}-1\right)^{m} \Bigg|_{x=0}
\end{align}
As a result, we find that the denominator may be evaluated to obtain
\begin{align}
&\sum_{s=0}^{N-m} s! \left\{ \begin{array}{c} k_{AB}\\s\end{array}\right\} \left(\begin{array}{c}N-m\\s\end{array}\right) = (N-m)^{k_{AB}}
\end{align}
so that the probability distribution takes the simpler form
\begin{align}
P(s) = \frac{s!}{(N-m)^{k_{AB}}}\left\{ \begin{array}{c} k_{AB}\\s\end{array}\right\} \left(\begin{array}{c} N-m\\s \end{array}\right)
\end{align}

To evaluate the average number of sites $N_{AB}$ that are reached in the $B$ sub-system we note that using Eq. (\ref{eq:StirlingS2_identity}), we obtain that
\begin{align}
&\frac{1}{(N-m)^{k_{AB}}}\sum_{s=0}^{N-m} s! \left\{ \begin{array}{c} k_{AB}\\s\end{array}\right\} \left(\begin{array}{c}N-m\\s\end{array}\right)s = (N-m)\left[1 - \left(1-\frac{1}{N-m}\right)^{k_{AB}}\right]
\end{align}
Let $q \equiv m/N$, and $k_{AB} = \langle k\rangle \,m (1-q)$, where $\langle k\rangle$ is the average degree of a node.  Then, in the limit that $N\rightarrow\infty$, with $q$ finite, we find
\begin{align}
\frac{N_{A,B}}{N} \longrightarrow (1 - q)\left[ 1- e^{-\langle k\rangle q}\right]
\end{align}
The entanglement entropy is bounded from above by $\min[N_{AB},N_{BA}]$, which then gives the result that 
\begin{align}
S_{A} \le q\left[ 1- e^{-\langle k\rangle (1-q)}\right] = q\left[ 1- e^{-(1 - e^{-\Gamma_{m}t})(1-q)/2g}\right]
\end{align}
when $q\le 1/2$.  %We observe that $d(t) = \langle k \rangle + 1$ is the average stabilizer size at time $t$, which is given by

\end{document}